\documentclass[12pt]{article}
\usepackage{amsfonts,amsmath}
 \usepackage{amssymb}
 \usepackage[hypertex] {hyperref}

\makeatletter \@addtoreset{equation}{section} \makeatother

\def\tga{{\tilde{\alpha}}}

\newcommand{\be}{\begin{equation}}
\newcommand{\ee}{\end{equation}}
\newcommand{\bee}{\begin{eqnarray}}
\newcommand{\beee}{\begin{array}}
\newcommand{\eee}{\end{eqnarray}}
\newcommand{\eeee}{\end{array}}

\newcommand{\un}{{\underline{n}}}
\newcommand{\um}{{\underline{m}}}

\newcommand{\cc}{{\cal C}}

\newcommand{\ga}{\alpha}
\newcommand{\pa}{{\dot{\ga}}}
\newcommand{\pb}{{\dot{\gb}}}

\newcommand{\gb}{\beta}
\newcommand{\gga}{\gamma}
\newcommand{\gla}{\lambda}

\newcommand{\M}{{\cal M}}

\newcommand{\E}{{\cal E}}

\newcommand{\W}{{\cal W}}
\newcommand{\K}{{\cal K}}
\newcommand{\F}{{\cal F}}

\newcommand{\Hh}{{\cal H}}

\newcommand{\ie}{{\it i.e.,} }
\newcommand{\ls}{\!\!\!\!\!\!}
\def\ck{{\mathcal K}}

\newcommand{\gl}{\lambda}

\newcommand{\gs}{\sigma}

\newcommand{\go}{\omega}

\newcommand{\by}{{\bar{y}}}

\newcommand{\q}{\,,\qquad}

\newcommand{\dga}{{\dot{\alpha}}}
\newcommand{\dgb}{{\dot{\beta}}}

\newcommand{\mR}{{\mathbb R}}

\newcommand{\nn}{\nonumber}

\newcommand{\half}{\frac{1}{2}}

\newcommand{\p}{\partial}
\newcommand{\D}{{\cal D}}

\newcommand{\f}{\frac}

\newcommand{\R}{{\cal R}}

\newcommand{\U}{\Upsilon}
\newcommand{\ups}{\upsilon}
\newcommand{\bu}{\bar{\upsilon}}

\def\a{\alpha}
\def\tga{{\tilde{\alpha}}}
\def\b{\beta}

\def\n{\um}

\def\o{\omega}

\def\ck{{\cal K}}
\def\cl{{\cal L}}

\def\cc{{\cal C}}

\def\ce{{\cal E}}

\newcommand{\Z}{{{\cal Z}}}
\newcommand{\gY}{{\mathcal{Y}}}

\newcommand{\hsum}{\subset{\ls +}}
\newcommand{\uA}{{\underline A}}
\newcommand{\uB}{{\underline B}}
\newcommand{\xx}{{\bf x}}
\newcommand{\ty}{\hat{y}}
\def\Di{${\mathcal A}$}
\def\Ne{${\mathcal B}$}

\begin{document}

\begin{flushright}
{\small FIAN/TD/06-12}
\end{flushright}
\vspace{1.7 cm}

\begin{center}
{\large\bf Holography, Unfolding and  Higher-Spin Theory}

\vspace{1 cm}

{\bf  M.A.~Vasiliev}\\
\vspace{0.5 cm}
{\it
 I.E. Tamm Department of Theoretical Physics, Lebedev Physical Institute,\\
Leninsky prospect 53, 119991, Moscow, Russia}

\end{center}

\vspace{0.4 cm}

\begin{abstract}
\noindent
Holographic duality is argued to relate classes of models
that have equivalent unfolded formulation, hence exhibiting
different space-time visualizations for the same theory. This
general phenomenon is illustrated by the $AdS_4$ higher-spin gauge
theory shown to be dual to the theory of $3d$ conformal currents of
all spins interacting with $3d$ conformal higher-spin fields of
Chern-Simons type. Generally, the resulting $3d$ boundary conformal theory is
nonlinear, providing an interacting version of the $3d$ boundary
sigma model conjectured by Klebanov and Polyakov to be dual to the
$AdS_4$ HS theory in the large $N$ limit. Being a gauge theory it
escapes the conditions of the theorem of Maldacena and Zhiboedov,
which force a $3d$ boundary conformal theory to be free.
Two reductions of particular higher-spin gauge theories where boundary
higher-spin gauge fields decouple from the currents and which have
free boundary duals are identified. Higher-spin holographic
duality is also discussed for the cases of $AdS_3/CFT_2$
and duality between higher-spin theories and nonrelativistic quantum
mechanics. In the latter case it is shown in particular that ($dS$)
$AdS$ geometry in the higher-spin setup is dual to the (inverted)
harmonic potential in the quantum-mechanical setup.

\end{abstract}

\newpage
\tableofcontents

\newpage

\section{Introduction}
\label{intro}
Higher-spin (HS) gauge theories describe interactions of massless
fields of all spins. First example of fully nonlinear HS theory
was  given in the $4d$ case in \cite{con}, while its
modern formulation was worked out  in \cite{more} (see \cite{Vasiliev:1999ba}
for a review). A specific property of HS gauge theories is that
consistent interactions of propagating massless fields require
a curved background  which provides a length
scale in HS  interactions that contain higher derivatives.
$(A)dS$ is the most symmetric curved
background compatible with HS interactions. The $AdS_4$ HS
model is the simplest nontrivial in the sense that $d=4$ is the
lowest dimension where HS massless fields propagate. After the
$AdS/CFT$ correspondence conjecture was put forward in
\cite{Maldacena:1997re,Gubser:1998bc,Witten:1998qj}, the fact that
HS theories are most naturally formulated in the $AdS$ background
was conjectured  to play a role in the context of
$AdS/CFT$ correspondence \cite{Sundborg:2000wp,WJ,BHS,
Mikhailov:2002bp,Sezgin:2002rt}.
This expectation conforms the fundamental
result of Flato and Fronsdal \cite{FF} on the relation between
tensor products of $3d$ conformal fields (singletons) and
infinite towers of $4d$ massless fields that appear in HS theories.

In the important
work of Klebanov and Polyakov \cite{Klebanov:2002ja} it was
argued that the HS gauge theory of \cite{more} should be
dual to the $3d$ $O(N)$ sigma model in the $N\to \infty$ limit.
The Klebanov-Polyakov conjecture was checked
by Giombi and Yin in \cite{Giombi:2009wh,Giombi:2010vg}
where it was shown in particular how the bulk computation in
HS gauge theory reproduces at least some
of conformal correlators in the free $3d$ theory.
(For related computations in free HS theory see also
\cite{Metsaev:2008fs,Metsaev:2009ym,Bekaert:2010ky,Joung:2011xb}.)
Recently, Maldacena and Zhiboedov \cite{Maldacena:2011jn}
addressed the question on restrictions imposed on a boundary $3d$
conformal theory by HS conformal symmetries. Assuming very general
conditions on conformal theory which included unitarity,
locality and conformal operator product algebra they were
able to show that a conformal HS theory, that possesses
a HS conserved current, should be free. This conclusion
seemingly suggests that any $AdS_4$ HS theory should be
equivalent to a free boundary theory at least in the most
symmetric vacuum.

The primary motivation for this work was to analyze directly a $3d$ dual of
the $AdS_4$ HS theory by means of the unfolded dynamics approach
which makes this analysis straightforward, describing evolution
with respect to different coordinates as independent mutually
commuting flows. This property allows us to obtain $3d$
field equations from $4d$ HS equations simply by reducing four
space-time coordinates  of $AdS_4$ to three,
relating directly two seemingly different theories in a way
anticipated from $AdS/CFT$ correspondence.

An important ingredient of the duality is that unfolded equations
for $3d$ conserved currents  result from the $3d$ reduction of $4d$
unfolded massless equations. The key observation is based on
the interpretation of currents as rank-two fields within the
approach of \cite{tens2} where it was shown that conserved currents,
built from the free fields described by $ C(Y|X) $ where $Y$ are
auxiliary spinor (twistor) variables while $X$ describe space-time
coordinates, are described by the fields $ J(Y^i|X) $ with the
doubled number of spinor variables $Y^i$, $i=1,2$. In particular,
free $3d$ massless fields are described by functions of
two-component spinors $y^\ga$ and space-time coordinates
$x^{\ga\gb}=x^{\gb\ga}$, where $\ga\,,\,\gb=1,2$, while $3d$
conserved currents $J(y^i|x)$ depend on a pair of spinors $y^i_\ga$.
On the other hand, a $4d$ massless field  is described by a field
$C(Y|X)$ where $Y=(y^\ga\,, \bar y^\dga)$ are complex two-component
spinors and $X=x^{\ga\dga}$ encode
 four real coordinates in the form of $2\times 2$ Hermitian
 matrices. It suffices to modify hermiticity conditions
 for two-component spinors to identify
 $4d$ massless fields with $3d$ conformal conserved currents.
The pullback of the known nonlinear $4d$ massless field equations
 to a $3d$ subspace $\Sigma^3\in AdS_4$ gives nonlinear $3d$
 equations which describe  interactions of conformal
 current fields with $3d$ conformal HS gauge fields. Precise
 identification only requires
an appropriate  change of the reality conditions and
transition to the conformal frame where  conformal symmetries
are manifest.

In this setup  holography takes place
for generic $3d$ surface $\Sigma$. However, the map from
the $AdS$ frame to the conformal frame, where conformal symmetries
are properly realized, turns out to be nonlocal for general $\Sigma$.
In the unfolded formulation of HS theory,    this map  has
simple meaning in terms of noncommutative twistor variables $Y$,
describing transition from the Weyl star product in the bulk theory to the
normal ordered one in the conformal frame. However,
in space-time terms  it may look obscure for general $\Sigma$.
Remarkably, in the limit where $\Sigma$ is $AdS_4$ infinity, the
correspondence between the two frames becomes local and very simple,
directly identifying $4d$ massless fields with (sources for) $3d$
currents in accordance with the original $AdS/CFT$ prescription of
\cite{Gubser:1998bc,Witten:1998qj}.

The conclusion  that nonlinear $AdS_4$ HS gauge theory is dual to
a nonlinear $3d$ theory does not contradict to Maldacena-Zhiboedov
theorem \cite{Maldacena:2011jn} because the boundary theory turns
out to be a
gauge theory with currents interacting through $3d$ Chern-Simons
conformal HS gauge fields. As such it escapes at least one of  the
assumptions of unitarity, locality and/or conformal invariance.
Indeed, gauge degrees of freedom correspond to null states,
while a gauge fixing procedure breaks covariance and/or locality.
There are, however, two special configurations for $AdS_4$ HS fields
that correspond to Dirichlet and Neumann boundary conditions in
the so-called $A$ and $B$ models,  where the $3d$
superconformal HS gauge fields decouple from currents. These
correspond to the free bosonic and fermionic boundary theories in
accordance with   Klebanov-Polyakov \cite{Klebanov:2002ja}
and Sezgin-Sundell \cite{Sezgin:2003pt} conjectures as well as
Maldacena-Zhiboedov theorem \cite{Maldacena:2011jn}.

Analysis of HS holography within unfolded dynamics
shows that phenomenon of holographic duality is absolutely
general, taking place in any $AdS$ theory. To make the
correspondence manifest, the theory in question has to be
reformulated  in the unfolded form. This makes the correspondence
to large extent tautological via reduction of space-time coordinates
in the unfolded theory. More precisely, the unfolding procedure
effectively reformulates a theory in terms of  appropriate
(generalized) twistor variables $Y$ rather than directly in
terms of space-time coordinates $X$.
In this setup, holography relates theories in different space-times
${\mathbf M}$ (coordinates $X$), that have the same description in the
twistor space ${\mathbf T}$
(coordinates $Y$). Reformulation of a theory in the unfolded form
makes it
straightforward to identify its holographic duals by choosing
different space-times ${\mathbf M}$ for the same twistor model.

In the general setup of this paper, the large $N$
parameter does not play any significant role. In HS theories it can be related
to the HS coupling constant so that in the large $N$ limit it
brings the boundary theory to the free field limit. However,
interactions with boundary HS gauge fields should be taken into
account in the analysis of subleading $1/N$ corrections.

Although in this paper we mainly discuss  the correspondence at the level
of field
equations because the appropriate action principle for HS gauge
theories remains unknown (for  some interesting conjectures see
however \cite{Boulanger:2011dd,Sezgin:2011hq}) it should be
stressed that general nature of holographic duality extends to
the action level.
Here we follow the analogy with the properties of conserved
currents discussed in \cite{gelcur} where
conserved charges in HS theory were represented in the form
\be
\label{Qc}
q=\int_\sigma \Omega (Y|X)\,,
\ee
where $\Omega(Y|X)$ is a closed $p$-form in some ``correspondence"
space ${\mathbf C}$ with local coordinates $Y$ (twistor space $\mathbf{T}$)
and $X$ (space-time ${\mathbf M}$). The charge $q$  is
independent of local variations of  $\sigma$ in ${\mathbf C}$
while $dim\, {\mathbf C}$ may be much larger than $p$.
For example, in the case of \cite{gelcur},  $q$
could be represented either in the standard form of a space
integral or as an integral in $\mathbf{T}$.

Since, as shown in \cite{act}, in the unfolded dynamics approach
the concepts of conserved charge and action  are  similar,
analogous representation is  anticipated to
be reachable for  HS actions. If so, the
generating functional in the bulk theory
\be
\label{Z}
Z_{bulk}(\phi_{boun}) = \int \exp i S(\phi (Y|X))
\ee
with  boundary conditions $\phi_{boun}$ for the bulk
dynamical variables $\phi (Y|X)$ can be represented in
many equivalent ways. If the integration is over space-time
$M_{bulk}$ this leads to the $AdS/CFT$ prescription.
If it is over some cycle in the twistor space $\mathbf{T}$
or in $M_{bound}\times \mathbf{T}$ this
provides an independent definition of the boundary theory.
(Note that $M_{bound}$ itself  has too small dimension
to support the integral).

An interesting output of the analysis of this paper is that,
at least for the  HS models in question, it is most
convenient to formulate them in the doubled $AdS$ space where the original
$AdS$ boundary
$z=0$ is identified with the invariant surface for the reflection automorphism
$P$ that maps one copy of $AdS$ to another via reflection $P (z)=-z$. As a result,
no boundary conditions at $z=0$ should be imposed to define the action
as the integral over the doubled $AdS$ space-time.
In this setup, holographic duality  relates a bulk theory in the doubled
$AdS$ space with the ``boundary theory" where all possible types of boundary fields
$\phi_{bound}(\xx)$ contribute. Values of  $\phi_{bound}(\xx)$ at  $z=0$ determine
all fields in the (doubled) bulk and hence values of the respective action
functionals $S(\phi_{bound})$. We believe that this doubling trick should have
a wide area of applicability  in HS theories and beyond.

The paper is organized as follows. In Section \ref{Unfolded
dynamics}, we recall relevant elements of the unfolded dynamics
approach. Interpretation of holography in terms of unfolded dynamics
is discussed in Section \ref{Duality}. Unfolded  equations for
massless fields in $AdS_4$ are recalled in Section \ref{free} while
unfolded formulation for $3d$ conserved currents is summarized in
Section \ref{currents}. In Section \ref{Nonlinear Higher-Spin
Equations}, we recall nonlinear HS equations in $AdS_4$ also
discussing in Section \ref{ferext}  their  extension with
spinor coordinates. In Section \ref{ads4cft3}, holographic duality
between HS theory in $AdS_4$ and $3d$ conformal HS theory is
discussed in general terms. Its detailed analysis in terms of
familiar Poincar{\'e} coordinates is presented in Section
\ref{cinf}. General structure of nonlinear $3d$ conformal HS theory
is discussed in Section \ref{n3dconf}. Boundary conditions, the construction
of the doubled $AdS$ bulk space and reductions
of particular HS theories, associated with free boundary theories, are
considered in Section \ref{bound}. $AdS_3/CFT_2$ HS
correspondence is briefly discussed in Section \ref{32}. Duality
between HS gauge theories and nonrelativistic quantum mechanics is
considered in Section \ref{qm}.
 Aspects of the off-shell extension of
on-shell unfolded theories are considered in Section \ref{actions}.
Conclusions and perspectives are presented in Section \ref{conc}.

\section{Unfolded dynamics}
\label{Unfolded dynamics}

\subsection{Unfolded equations}
\label{fda}

Unfolded formulation is a multidimensional coordinate independent
 generalization of the first--order formulation
\be
\label{d1}
dt \f{\p}{\p t}{q}^\ga = G^\ga (q)\,,\qquad G^\ga (q)= e F^\ga (q)
\ee
available for any system of ordinary differential
equations by adding auxiliary variables associated with
higher derivatives of the dynamical variables of the original
system. Here $e$ is a einbein 1-form
that can be identified with $dt$ because $1d$
geometry is  flat.

Let $M^d$ be a $d$-dimensional space-time manifold
with coordinates $x^\un$ ($\un = 0,1,\ldots d-1$).
By unfolded formulation of a linear or nonlinear
system of partial differential equations (PDE)
in $M^d$ we mean its reformulation in the first-order form
\cite{Ann}
\be
\label{unf} dW^\Omega (x)= G^\Omega (W(x))\,,
\ee
where $ d=dx^\un  \frac{\p}{\p x^\un}\, $
is the exterior  differential in $M^d$, $W^\Omega(x)$
is a set of degree $p_\Omega$ differential forms
and $G^\Omega (W)$ is some degree $p_\Omega +1$
function of $W^\Lambda$
\be
G^\Omega (W) =
\sum_{n=1}^\infty f^\Omega{}_{\Lambda_1\ldots \Lambda_n} W^{\Lambda_1}\wedge \ldots
\wedge W^{\Lambda_n}\,,
\ee
where the coefficients $f^\Omega{}_{\Lambda_1\ldots
\Lambda_n}$ satisfy the (anti)symmetry condition
\be
f^\Omega{}_{\Lambda_1\ldots\Lambda_k \Lambda_l \ldots \Lambda_n} =
(-1)^{p_{\Lambda_k}p_{\Lambda_l}}
f^\Omega{}_{\Lambda_1\ldots\Lambda_l\Lambda_k \ldots  \Lambda_n} \,
\ee
(extension to the case
with additional boson-fermion grading is straightforward)
and $G^\Omega(W)$ satisfies the  condition
\be \label{BI} G^\Lambda (W)\wedge \f{\p
G^\Omega (W)} {\p W^\Lambda}  =0\,
\ee
equivalent to the
following generalized Jacobi identity on the structure coefficients
\be \label{jid}
\sum_{n=0}^{m} (n+1) f^\Phi{}_{[\Lambda_1 \ldots
\Lambda_{m-n}}  f^\Omega{}_{\Phi\Lambda_{m-n+1} \ldots \Lambda_m\}} =0\,,
\ee
where
the brackets $[\,\}$ denote an appropriate (anti)symmetrization of
all indices $\Lambda_i$. Strictly speaking,  generalized Jacobi
identities (\ref{jid}) have to be satisfied  at $p_{\Omega} < d$
since any $(d+1)$-form  in ${ M}^d$ is zero. Given solution of
(\ref{jid}) it defines a free differential algebra
\cite{Sullivan,FDA,FDA1,FDA2}. We  call a free differential algebra {\it
universal} \cite{Bekaert:2005vh,act}
if the generalized Jacobi identity holds  independently of
a particular value of space-time dimension. All HS free
differential algebras relevant to HS theories including those
discussed in this paper are universal. For example,
the  $1d$ system (\ref{d1})  is universal.
Here the condition (\ref{BI}) trivializes because $e\wedge e=0$,
{\it i.e.,} any function $F^\ga (q)$ is allowed.
The generalized Jacobi identity is obeyed  for any number
of coordinates of the ambient space (i.e., $dx^\um$),
since  $e = dx^\um e_\um$  carries no fiber indices.

Condition (\ref{BI}),  which can equivalently be rewritten as
\be
\label{qdif}
Q^2 =0\,,\qquad Q= G^\Omega (W)  \f{\p}{\p W^\Omega}\,,
\ee
guarantees formal consistency of the unfolded system (\ref{unf})
which can now be put into the form
\be
\label{unf1}
d F(W(x)) =  Q (F(W(x))
\ee
with $d^2=0$ for all $F(W)$.
Unfolded equations in the form (\ref{unf1}) are
analogous to $1d$ Hamiltonian equations.

Equation (\ref{unf}) is invariant under the gauge transformation
\be \label{delw} \delta W^\Omega = d \varepsilon^\Omega +\varepsilon^\Lambda
\frac{\p G^\Omega (W) }{\p W^\Lambda}\,,
\ee
where the gauge parameter
$\varepsilon^\Omega (x) $ is a $(p_\Omega -1)$-form.  ($0$-forms
among $W^\Omega$ do not have associated gauge parameters.)

\subsection{Examples}

\subsubsection{Vacuum}
\label{vacuum} Important example of unfolded equations is provided
by Maurer-Cartan equations. Let $h$ be a Lie algebra with some basis
$\{T_\a\}$. Let $w=w^\a T_\a$ be a $h$-valued 1-form. Setting $G
(\,w)=-w\wedge w\equiv -\frac{1}{2} w^\a \wedge w^\b [T_\a , T_\b]$,
Eq.~(\ref{unf}) with $W=w$ becomes the flatness condition \be
\label{0cur} dw+w\wedge w=0\,. \ee Eq.~(\ref{BI}) amounts to the
Jacobi identity for  $h$. Eq.~(\ref{delw}) gives the usual gauge
transformation \be \delta w =D_0 \varepsilon := d
+[w\,,\varepsilon]\,. \ee

Usually, the zero-curvature equations (\ref{0cur})  describe
background geometry in a coordinate independent way.
For example, let $h$ be Poincar{\'e} algebra with the gauge fields
\be
w (x)=e^n (x) P_n
+\o^{nm} (x) L_{nm}\,,
\ee
where $P_n$ and $L_{nm}$ are generators of
translations and Lorentz transformations. The  gauge
fields $e^n (x)$ and $\o^{nm}(x)$ are identified with the frame 1-form
and Lorentz connection, respectively
(fiber Lorentz vector indices $m,n\ldots$ and base indices $\um,\un\ldots $
run from 0 to $d-1$
and are raised and lowered by the flat Minkowski metric).
It is well known that the zero-curvature
condition (\ref{0cur}) for the Poincar{\'e} algebra amounts to the
zero-torsion condition
\be
\label{tor}
R^n=de^n +\o^n{}_m \wedge e^m = 0\,,
\ee
which expresses $\o^{nm}(x)$ in terms of derivatives of
the frame field, and the condition that Riemann tensor is zero
\be
\label{riem}
R^{mn}=d\o^{mn}
+\o^m{}_k \wedge \o^{kn}=0\,,
\ee
which implies flat Minkowski  geometry.
As a result, at the condition that the matrix
$e_{\un}{}^n (x)$ is nondegenerate, the zero-curvature
condition (\ref{0cur}) for the Poincar{\' e} algebra gives
coordinate independent description of Minkowski space-time.
Choosing a different Lie algebra $h$ one can describe
different background like, e.g., (anti-) de Sitter.

\subsubsection{Linear fluctuations}
\label{sca}
If the set $W^\Omega$ contains some $p$-forms denoted by $\cc^i$
({\it e.g.} $0$-forms) and  $G^i(W)$ is linear
both in $w$ and in $\cc^i$
\be
\label{lin}
 G^i =- w^\a(T_\a)^i {}_j \wedge \cc^j\,,
\ee
 relation
(\ref{BI}) implies that $(T_\a)^i {}_j$ form some
representation $T$ of $h$, acting in a space $V$ where  $\cc^i$
is valued. The corresponding equation (\ref{unf}) is a
covariant constancy condition
\be \label{covc} D_w \cc^i=0
\ee
 with $D_w\equiv d+w$ being the covariant derivative
in the $h$-module $V$. For $G^i(w,\cc^i)$ multilinear
in the background connections $w$ but still linear in
dynamical fields $\cc^j$, unfolded equations can be interpreted
in terms of Chevalley-Eilenberg cohomology of $h$ with
coefficients in the infinite-dimensional modules
carried by differential forms of different degrees among
$\cc^j$ \cite{333}.

As an illustration, consider unfolded formulation of a scalar field.
Following \cite{sigma} we introduce  the infinite set of
$0$-forms $C_{m_1\ldots m_n}(x)$ ($n=0,1,2,\ldots$), which are
totally symmetric tensors
$C_{m_1\ldots
m_n}=C_{(m_1\ldots m_n)}\,.
$
The off-shell unfolded equations are
 \be \label{un0} d C_{m_1\ldots m_n } =e^k C_{m_1
\ldots m_n k}\qquad (n=0,1,\ldots)\,,
\ee
where we use Cartesian coordinates with $D^L = d$.
This system is formally consistent because application of
$d$ to the both sides of (\ref{un0}) gives zero by
 $e^k \wedge e^l=-e^l \wedge e^k$. Hence, the space $V$ of $0$-forms
$C_{m_1 \ldots m_n}$ forms some (infinite-dimensional)
$iso(d-1,1)$--module. (Strictly speaking,
one has to check that the equation
is consistent for any flat $iso(d-1,1)$ connection. It is
not hard to see  that this is indeed true.)

 Let  the scalar field $C (x)$ be identified with $C_{m_1
\ldots m_n}(x)$ at $n=0$. The first two equations of the system
(\ref{un0}) read
\be
\label{fieq}
\partial_\un C =C_\un \,,\qquad
\partial_\un C_\um= C_{\um\un}\,,
\ee
where we have identified the world and tangent indices via
$e_\um^m=\delta_\um^m$. The first of these equations  tells
us that $C_\un$ is the first derivatives of $C$. The second one
identifies $C_{\un\um}$ with the second derivative of $C$.
 All other
equations in (\ref{un0}) express highest tensors in terms of the
higher-order derivatives \be \label{hder} C_{\um_1 \ldots \um_n}=
\partial_{\um_1} \ldots \partial_{\um_n}C\,.
\ee
{}From  Eq.(\ref{hder}) it
is clear that the  0-forms $C_{\un_1 \ldots \un_n}$  describe
all derivatives of the dynamical field $C(x)$ including
$C(x)$ itself. The system (\ref{un0}) is
off-shell: it is equivalent to an infinite set of constraints,
imposing no field equations on the dynamical field
$C$.

To put the system on shell we impose an additional condition that
all $C_{m_1\ldots m_n}(x)$ are traceless
\be
\label{trace}
C^k{}_{k m_3\ldots m_n}(x)=0\,.
\ee
This condition preserves consistency and  puts
the system on shell by virtue of Eq.~(\ref{fieq}).

\subsection{Global symmetries}
\label{globs}
Maximally symmetric vacua of dynamical systems are described by  vacuum
connections that satisfy the flatness condition (\ref{0cur})
for some Lie algebra $h$.
Choosing some vacuum connection $w_0 (x)$, global
symmetry transformations that leave $w_0$ invariant
are described by the parameters $\varepsilon_{gl}(x)$ that
satisfy
\be
\label{glo0}
D_0 \varepsilon_{gl}=0\,.
\ee
Clearly, in the topologically trivial situation
this equation has $dim\, h$ independent solutions.

This simple observation immediately uncovers maximal symmetries
of the linear unfolded equations of the form (\ref{covc}).
Indeed, an $h$-module $V$ can be treated as the $l^{max}(V)$-module
where $l^{max}(V)$ is the Lie algebra of commutators of
$End\,V$. Indeed, since $h\in l^{max}(V)$, any flat connection
$w_0$ of $h$ can be interpreted as a flat connection of
$l^{max}(V)$. Hence, $l^{max}(V)$ is the maximal
symmetry of the linear unfolded equations with dynamical fields
valued in $V$. As a result, via identification of $V$, unfolding
of a dynamical system makes all its symmetries manifest.

Let $W_0^\Omega $ be some solution of the unfolded system
(\ref{unf}), may be containing some nonzero $p_\Omega$-forms with
$p_\Omega\neq 1$. According to (\ref{delw}),
 symmetries of this solution are described by
 the symmetry parameters $\epsilon_{gl}^\Omega (x)$ that satisfy
\be \label{glo} d \varepsilon_{gl}^\Omega +\varepsilon_{gl}^\Lambda
\frac{\p G^\Omega (W) }{\p W^\Lambda}\Big |_{W=W_0}=0\,.
\ee
Since equations (\ref{glo}) that contain
$d \varepsilon_{gl}^\Omega$ are
consistent as a consequence of the original unfolded equations
they can be solved locally in terms of $\varepsilon_{gl}^\Omega(x_0)$
at any space-time point $x_0$. Naively, it looks like this
gives as many global symmetries as parameters
$\varepsilon_{gl}^\Omega$. However, this is not
the case because the 0-form part  of Eq.~(\ref{glo})
may  impose constraints on
 $\varepsilon_{gl}^\Omega(x)$
\be \label{cons} {}^0\varepsilon_{gl}^\Lambda
\frac{\p G^\Omega (W) }{\p\, {}^1 W^\Lambda}\Big |_{ W={}^0 W_0}=0\,,
\ee
where ${}^p W^\Omega $ denotes $p$-forms among $W^\Omega$.
This  implies that nontrivial global symmetries
should leave invariant vacuum values of 0-forms in the
system. This restriction is very strong since,
in the unfolded dynamics approach, nontrivial curvatures like
Weyl tensor and its HS analogues are described by 0-forms.
Most symmetric vacua are  associated with those solutions
where all 0-forms are zero or central (\ie singlet with respect
to all 1-form connections).

\subsection{Dynamical content}
\label{dyncon}

General situation can be illustrated by the simple $1d$
example. First-order ordinary differential
equations have the following structure
\be
\label{eqex}
 \f{\p}{\p t}{q}^\tga_i =a_i^\tga {q}^\tga_{i+1} +\ldots\,,
\qquad i=0,1,2,\ldots\,,
\ee
where $a_i^\tga$ are some coefficients while ellipsis denote
nonlinear corrections. If all $a^\tga_i$ are nonzero,
Eqs.~(\ref{eqex}) treated perturbatively describe
an infinite set of constraints that express all ${q}^\tga_{i+1}$ via
derivatives of  ${q}^\tga_{0}$.
If some coefficient $a_j^\tga$ vanishes, this implies a
nontrivial differential equation on ${q}^\tga_{0}$,
which is of order $j$ at the linearized level.
These two options are analogous, respectively,
to the off-shell and on-shell cases in the scalar field example
of Section \ref{sca}. Indeed, using that $1d$ traceless
tensors in one dimension are  all zero except for
$C$ and  $C_n$,  the on-shell system (\ref{un0}),
(\ref{trace}) at $d=1$ implies $\f{\p^2}{\p t\p t } C$=0.

In the first-order formulation
(in particular, in the Hamiltonian formalism) the initial data
problem is set in terms of values of all variables $q$
at given time $t_0$.
In the general case of $d>1$ these properties have clear analogues.
Nontrivial dynamical fields ({\it i.e.,} those that are different from
{\it auxiliary fields} expressed via derivatives of the
dynamical fields), gauge symmetries and true differential field
equations, are classified in terms of the so-called $\sigma_-$
cohomology \cite{sigma}
 that roughly speaking controls zeros
among the coefficients analogous to $a^\tga_i$ of the linearized
equations.
The $\sigma_-$ cohomology is a perturbative
concept that emerges in the linearized analysis with
\be
W^\Omega (x) = W^\Omega_0 (x) + W_1^\Omega (x)\,,
\ee
where $W^\Omega_0 (x)$ is a particular solution
of (\ref{unf}) and $W_1^\Omega (x)$ is treated as a perturbation.
$W^\Omega_0 (x)$ is nonzero in a field-theoretical
system because, as explained in Section \ref{vacuum}, it
should contain a background gravitational field. Linearized
equations (\ref{unf})
\be
dW_1^\Omega (x) = W_1^\Lambda (x) \f{\delta G^\Omega}{\delta W^\Lambda }
\Big |_{W=W_0}
\ee
can be rewritten in the form (\ref{covc})
\be
\label{d0}
D_0 W_1^\Omega (x)=0\,,
\ee
where $D_0$ is some differential that squares to zero to fulfill
the consistency condition (\ref{BI}),
\be
\label{d02}
D_0^2 =0\,.
\ee

Usually, a set of fields $W_1$ admits a grading $G$ with the
spectrum bounded from below, which  typically counts a
rank of a tensor.
Suppose that \be \label{d0s} D_0 = \D_0
+\sigma_- +\sigma_+\,, \ee where \be \label{gs-} [G\,,\sigma_- ] =
- \sigma_-\,,\qquad [G\,,\D_0 ]= 0\, \ee and $\sigma_+$ is a sum
of operators of positive grade.
From (\ref{d02}) it follows that
\be \sigma_-^2 = 0\,. \ee
Provided that $\sigma_-$ does not
differentiate $x^\un$, dynamical content of the
system in question is determined by cohomology of
 $\sigma_-$. Namely, as shown in \cite{sigma} (see also \cite{Bekaert:2005vh}),
 for
$p_\Omega$-forms $W^\Omega$ valued in a vector space $V$,
$H^{p+1} (\sigma_- ,V)$, $H^{p} (\sigma_- ,V)$ and $H^{p-1}
(\sigma_- ,V)$ describe, respectively, differential equations,
dynamical fields and differential gauge symmetries encoded by
equation (\ref{d0}). The case with $H^{p+1} (\sigma_- ,V)=0$ is
analogous to that of (\ref{eqex}) with all
coefficients $a^\tga_i$ different from zero.
Here no differential equations  on the dynamical
variables are imposed, \ie  equations (\ref{unf}) just encode
 constraints on auxiliary fields. Equations of this type
are referred to as {\it off-shell}.
(Let  us stress that this definition is true both for
linear and  non-linear cases: nonlinear equations are
off-shell if their linearization is off-shell.)
If $H^{p+1} (\sigma_- ,V)\neq 0$, unfolded equations (\ref{unf})
impose some differential equations on the dynamical fields.
Such systems are called {\it on-shell}.

Degrees of freedom, \ie variables that determine a (local)
solution of equations (\ref{unf}) modulo gauge ambiguity,
are represented by
values of all 0-forms $C^\phi(x_0)$ among $W^\Omega (x_0)$
at any given $x_0$  which is
analogous to $t_0$ of the $1d$ case. This means in particular
that to unfold a field-theoretical
system with infinite number of degrees of freedom, an infinite set
of 0-forms has to be introduced.
In the scalar field example this is  the
set of 0-forms $C_{n_1\ldots n_k}(x)$.

\subsection{Generalized twistor space}

Unfolded scalar field system is most conveniently described
in terms of generating functions  of auxiliary
variables $y^n$
\be
C(y|x) =\sum_{k=0}^\infty \f{1}{k!}
y^{n_1}\ldots y^{n_k} C_{n_1\ldots n_k}(x)
\,.
\ee
In the on-shell case, tracelessness of $C_{n_1\ldots n_k}(x)$
is equivalent to the condition that $C(y|x)$ is harmonic in
$y^n$
\be
\label{box}
\Box_y C(y|x)=0\,
\ee
while the unfolded equations (\ref{un0}) take the form
\be
\label{un0y}
(d_x - e^n \f{\p}{\p y^n}) C(y|x) =0\,.
\ee
In these terms
\be
G =y^n \f{\p}{\p y^n} \q D_0 =d \q \sigma_+ =0\q
\sigma_- = - e^n \f{\p}{\p y^n}\,.
\ee
In the off-shell case, $\sigma_-$ acts as exterior differential
on polynomials. Hence, the only nontrivial $\gs_-$--cohomology is
$H^0(\sigma_-)$.
This tells us that scalar $C(x)$ is the only dynamical field and
that it is not restricted by dynamical equations. In the on-shell
case, $H^1(\sigma_-)$ is one dimensional in agreement with the
fact that the on-shell unfolded system imposes only one equation
on the scalar field, namely Klein-Gordon equation, which
belongs to the following part of the unfolded equations
\be
e^\n_m \f{\p^2}{\p dx^\n \p y_m}
\Big (d_x -e^a \f{\p}{\p y^a}\Big ) C(y|x) \Big |_{y=0} =0\,,
\ee
where the vector field $e^\n_m$ is the inverse to the (co)vielbein 1-form $e_\n^m$.

In field-theoretical systems with  infinite degrees of
freedom the label $\Omega$, enumerating
differential forms in the unfolded equations
(\ref{unf}), usually refers to appropriate functional spaces, \ie
$W^\Omega(X)$ can be represented as a set  $W^i(Y|X)$
of functions of some auxiliary variables $Y$ ($i$ labels different
functions).

The variables $Y$  are analogous to the
twistor coordinates in the  twistor theory. Following
\cite{gvtw} we  interpret $Y$ as coordinates of a generalized
twistor space\footnote{In HS theories, $\mathbf{T}$
 may or may not have precise geometric
meaning of some twistor space in the twistor theory
\cite{Bast_East}.
Nevertheless, abusing terminology,  in this paper we
 call it twistor space.} $\mathbf{T}$.
$X$ are space-time coordinates of space $\mathbf{M}$.
Together,  $(Y,X)$  are interpreted as
coordinates of the  correspondence space $\mathbf{C}$.
More precisely, $\mathbf{C}$ is a fiber bundle with $\mathbf{M}$
as base manifold and $\mathbf{T}$ as fibers.

In this setup, the unfolded equation (\ref{unf}) encodes a
generalized Penrose transform \cite{penr,Bast_East}
expressed by the diagram\bee\label{diapen}
 \begin{picture}(200,80)( 0,26)
{\linethickness{.25mm}
\put(90,90){\vector( 1,-1){40}}%
\put(90,90){\vector( -1,-1){40}}%
\put(85,95)  {{    $\mathbf{C}$}}
\put(30,35)  {{   $\mathbf{M}$}}
\put(128,35)  {{   $\mathbf{T}\quad .$}}
\put(55,70)  {{  \large $\eta$}}
\put(108,70)  {{  \large $\nu$}}
 }
\end{picture}
\eee
Indeed, it reconstructs (locally) general
solution $W^i(Y|X)$ of the unfolded equations in terms
of arbitrary functions $W^i(Y|X_0)$ on the twistor space
$\mathbf{T}$. Via restriction of a $p$-form $W^i(Y|X)$ to dynamical fields
associated with $H^p(\sigma_-)$
this gives a solution of the dynamical equations associated
with the $H^{p+1}(\sigma_-)$ part of the unfolded equations.

As discussed in more detail in Section \ref{free},
in lower dimensions $d=3,4$, the on-shell condition (\ref{box})
is conveniently resolved in terms of unrestricted functions
of spinor (twistor) variables. This greatly simplifies analysis
of the respective unfolded on-shell systems.

As discussed  in Section \ref{Duality},
holographic duality relates differently looking dynamical systems
in different space-times (coordinates $X$) that
have the same twistor space. From this perspective
unfolded equations perform a generalized Penrose transform from
the same twistor space $\mathbf{T}$ to one or another space-time
$\mathbf{M}$.

\subsection{Properties}
\label{prop}

Unfolded formulation has a number of useful and important properties
discussed in some more detail, {\it e.g.}, in \cite{333}.
In particular, unfolded equations possess manifest gauge
and diffeomorphism invariance  due to exterior algebra formalism.
As such, they are perfectly suited for the study of gauge invariant
theories in the framework of gravity like, {\it e.g.}, HS gauge
theories. The following properties
of unfolded dynamics are most relevant to the analysis of holography.

The method is universal. Any dynamical system can
be unfolded. This is analogous to the fact
that any system of ordinary differential equations admits a
first-order formulation.

Indeed, let $w=e_0^a\,P_a+\frac12 \go_0^{ab}M_{ab}$ be a vacuum
connection valued in some space-time
symmetry algebra $h$. Let a field $C^{(0)}(X)$
satisfy some dynamical equations to be unfolded. Consider
first the case where $C^{(0)}(X)$ is a 0-form.
One starts by writing the equation
$D_0^L C^{(0)} \,=\,
e_0^a\,\,\, C_a^{(1)},
$
 where $D_0^L$ is the Lorentz
covariant  derivative and the field $C_a^{(1)}$ is auxiliary.
Next, one checks whether the original field equations for $C^{(0)}$
impose restrictions on the first derivatives of $C^{(0)}$.
A part of $D^L_{0\um} C^{(0)}$, and hence $C^{(1)}$, may vanish on
the mass-shell ({\it e.g.} for Dirac equation). The remaining
non-zero auxiliary
fields  $C^{(1)}$ parameterize on-mass-shell nontrivial components
of first derivatives. One proceeds by writing analogous equation
for the first-level auxiliary  fields
$ D_0^L C^{(1)}_a =  e_0^b\,\, C^{(2)}_{a,b}
$
where the new fields $C^{(2)}_{a,b}$ parameterize  second
derivatives of $C^{(0)}$. Again, one checks, taking into
account  Bianchi identities, which components of  the second
level fields $C^{(2)}_{a,b}$ remain nonzero if the
original equations of motion are satisfied.
This process continues indefinitely, leading to a chain of
equations on the chain of fields $C^{(m)}_{a_1,a_2,\ldots,a_m}$
($m\in\mathbb N$) parameterizing all  on-mass-shell nontrivial
derivatives of the original dynamical field.
If one starts with some gauge field, analogous analysis
determines a form of shift gauge
transformations that subtract extra field components
to be introduced to describe a system in terms of differential
forms. (For instance, local Lorentz symmetry in Cartan formulation
of gravity  appears this way as the shift symmetry that removes extra
components of the vielbein 1-form compared
to the metric tensor.) By construction, this
leads to a particular unfolded system. The correspondence
between $p\geq 1$ forms and gauge symmetries in the unfolded dynamics
approach uncovers  pattern of
local and global symmetries associated with a given gauge field.
In particular, the pattern of the linearized $4d$ HS algebras was
deduced this way in \cite{V1}. These results were then used in
\cite{FVA,OP1} to find infinite-dimensional non-Abelian HS
algebras that underly the nonlinear $4d$ HS theories.

In the topologically trivial situation, degrees of freedom are
carried by 0-forms  at any space-time point $X=X_0$. Indeed, by
virtue of
Poincar{\'e} lemma, unfolded equations express all exterior derivatives
in terms of the values of fields themselves modulo exact forms
that can be gauged away by the gauge transformation
(\ref{delw}). What is left is the ``constant part" of  0-forms.
In terms of  functions of  twistor variables
$Y$ like $C^i(Y|X)$ this means that dynamics is entirely encoded
by 0-forms on the twistor space, \ie $C^i(Y|X_0)$ at any $X_0$.

\section{Unfolding and holographic duality}
\label{Duality}
Unfolded formulation unifies various dual versions of one and the
same system. This concerns both duality between
systems in the same space-time and holographic duality
between theories in space-times of different dimensions.

In the former case, duality results from the ambiguity in which
fields are chosen  to be dynamical or auxiliary, the nomenclature
governed by the choice of the grading $G$ and $\sigma_-$. Different
gradings lead to different interpretations of the same unfolded
system in terms of different dynamical fields that satisfy seemingly
unrelated differential equations. The key point is that if two
dynamical systems give rise to the same unfolded system (more
precisely, belong to the same projective system \cite{333}), they
are equivalent.

Holographic duality rests on  the  striking feature
that a universal unfolded system may admit different
space-time interpretations. In some sense,  space-time
dependence in such systems is auxiliary as
was first noted in the context of HS dynamics in \cite{alg}.
True dynamics is hidden in the twistor sector of the
auxiliary variables $Y$.

Indeed, in a universal unfolded system dynamics is entirely encoded by
the function $G^\Omega(W)$ independently of the original space-time
picture. In particular, unfolded formulation
allows one to extend space-time without changing dynamics simply by
letting the differential $d$ and differential forms $W^\Phi$
to live in a larger space
\be\label{dext}
d=dX^n\f{\p}{\p X^n}  \rightarrow \tilde{d}=dX^n\f{\p}{\p X^n}
+d\hat X^{\hat{n}} \f{\p}{\p \hat{X}^{\hat{n}}}\q
dX^n W_n \to
dX^n W_n + d\hat{X}^{\hat{n}} \hat{W}_{\hat{n}}\,,
\ee
where $\hat{X}^{\hat{n}}$ are some additional coordinates.
For a universal unfolded system such  substitution
neither spoils consistency nor affects local dynamics
still determined by the 0-forms at any point of
(any) space-time. Indeed, the unfolded system
in the $X$ space remains a subsystem of that in the enlarged space
while additional equations reconstruct  dependence on
$\hat{X}^{\hat{n}}$ in terms of solutions of the original
system (of course, this consideration is local).

On the other hand, as emphasized in \cite{Mar},
the role of coordinates is that they help to visualize
physical local events via a  physical processes.
A particular space-time interpretation of a universal
unfolded system, e.g,
whether a system is on-shell or off-shell, depends not only on
$G^\Omega (W)$ but, in the first place, on a chosen  space-time $M^d$ and
vacuum solution $W_0 (X)$. Dynamical interpretation may
be different for different  space-times
because $\gs_-$ cohomology depends on space-time dimension
via rank of the vielbein
1-form. In particular, the on-shell HS
theory in $AdS_4$ will be shown in Section \ref{n3dconf} to be
dual to an off-shell $3d$ conformal system.  This
implies that the $3d$ ``dynamical" boundary fields are not restricted
by any differential field equations. Among other things this property
makes it possible to identify $3d$ dynamical fields with unrestricted boundary
values of the on-shell bulk fields.

Important point considered in more detail in Section \ref{actions}
is that, in unfolded formulation, a nontrivial conserved
charge or gauge invariant generating functional for correlators
is represented as integral of some $d$-closed form \cite{act}.
As a result, correlators in boundary conformal systems are represented by
integrals over a space larger than space-time
where conformal fields live. It can be either the bulk space
as in the standard $AdS/CFT$ treatment or
some twistor space. In any case, from this perspective
the nonlinear off-shell system at the boundary is represented
by a nonlinear on-shell system in a larger space\footnote{There
is a subtlety related to this point to be taken into account in the
holography context.
Perturbatively, every nonlinear off-shell system is equivalent to
some linear  system by virtue of a perturbatively
local nonlinear field redefinition. Indeed, a system is off-shell
in the case where the operator $\sigma_-$ is a kind of
nondegenerate. Writing schematically
$
R_n = dw_n +\sigma_- w_{n+1} +\ldots\,,
$
where ellipsis denote  lower-grade and/or nonlinear terms one
observes that the latter can all  be absorbed into nonlinear
field redefinitions
of $w_{k}$ with $k=1,2,\ldots$ modulo $\sigma_-$ exact
terms that are pure gauge with respect to some shift
 symmetries. Although such a field redefinition
 destroys the structure of unfolded equations,
 the resulting system has linear form. This should be
compared to on-shell nonlinear systems where
 not all nonlinear terms can be removed by such a
field redefinition. In particular, the l.h.s.s of field equations,
associated with the respective $\sigma_-$ cohomology,
cannot be linearized in a general nonlinear on-shell system.}.

To summarize, two unfolded systems in different space-times are
equivalent (dual) provided that they have the same unfolded form. Generally,
a most straightforward way to establish
holographic duality between two theories is to unfold both of
them to see  whether the operators $Q$ (\ref{qdif}) of their unfolded
formulations coincide. Other way around, given unfolded system
generates a class of holographically dual theories in different
dimensions. It should be stressed that, being simple in terms
of unfolded dynamics and the corresponding twistor space
$\mathbf{T}$, holographic duality in usual space-time
terms may be very complicated requiring solution of at least
one of the two unfolded systems which is equivalent to a
nonlinear integral transform.

\section{Free massless fields in $AdS_4$}
\label{free}

In this section, we remind the reader
unfolded formulation of free massless fields of all spins in $AdS_4$
obtained originally in \cite{Ann}. It is based on
the frame-like approach to HS gauge fields \cite{Vasiliev:1980as,V1}
where a $4d$ spin $s\geq 1$  massless field is described by the set of
1-forms
\be\label{1f} \omega_{\ga_1\ldots
\ga_k,\pa_1\ldots\pa_l}= dx^\un \omega_{\un\,\ga_1\ldots
\ga_k,\pa_1\ldots\pa_l}\q k+l=2(s-1)\,,
\ee
where $\ga,\gb\ldots = 1,2$ and $\pa,\pb\ldots = 1,2$
are two-component spinor indices.
 The HS gauge fields are totally symmetric with respect to
 each type of spinor indices and obey the reality conditions
$\overline{\omega_{\ga_1\ldots \ga_k\,,\pb_1\ldots\pb_l}}=
\omega_{\gb_1\ldots\gb_l\,,\pa_1\ldots \pa_k}.
$ For a given $s$, this set is equivalent to the real
1-form $\omega_{A_1\ldots A_{2(s-1)}}$
symmetric in the  Majorana indices $A=1,2,3,4$. As such it
carries an irreducible module of $sp(4,{\mathbb R})\sim o(3,2)$.

$AdS_4$ is
described by the Lorentz connection $\omega^{\ga\gb}$,
$\overline{\omega}^{\pa\pb}$ and vierbein $e^{\ga\pa}$.
Altogether they form the $sp(4,{\mathbb R})$ connection $w^{AB}=w^{BA}$
that satisfies the $sp(4,{\mathbb R})$ zero-curvature conditions
\be
\label{ads}
R^{AB}=0 \,,\qquad R^{AB} = dw^{AB} +w^{AC}\wedge w_C{}^B\,,
\ee
where indices are raised and lowered by a $sp(4,\mR)$
invariant form $C_{AB}=-C_{BA}$
\be
\label{Cind}
A_B =A^A C_{AB}\q A^A = C^{AB} A_B\q C_{AC}C^{BC} = \delta_A^B\,.
\ee
In terms of Lorentz components
$
w^{AB} = (\go^{\ga\gb}, \overline{\go}^{\pa\pb}, \gla e^{\ga\pb},
\gla e^{\gb\pa} ),
$
where $\lambda^{-1}$ is the  $AdS_4$ radius,
the  $AdS_4$ equations (\ref{ads}) read as
\be
\label{adsfl}
R_{\ga\gb}=0\,,\quad \overline{R}_{\pa\pb}=0\,,
\quad R_{\ga\pa}=0\,,
\ee
where
\be
\label{nR}
R_{\alpha \gb}=d\omega_{\alpha \gb} +\omega_{\alpha}{}^\gamma
\wedge \omega_{\gb \gamma} +\lambda^2\, e_{\alpha}{}^{\dot{\delta}}
\wedge
e_{\gb \dot{\delta}}\,,
\ee
\be
\nn
\overline{R}_{{\pa} {\pb}}
=d\overline{\omega}_{{\pa}
{\pb}} +\overline{\omega}_{{\pa}}{}^{\dot{\gamma}}
\wedge \overline{\omega}_{{\pb} \dot{\gga}} +\lambda^2\,
e^\gamma{}_{{\pa}} \wedge e_{\gamma {\pb}}\,,
\ee
\begin{equation}
\label{nr}
R_{\alpha {\pb}} =de_{\alpha{\pb}} +\omega_\alpha{}^\gamma \wedge
e_{\gamma{\pb}} +\overline{\omega}_{{\pb}}{}^{\dot{\delta}}
\wedge e_{\alpha\dot{\delta}}\,.
\end{equation}
(Two-component indices are raised and lowered as in (\ref{Cind})
with $\epsilon_{\ga\gb}$ or $\epsilon_{\pa\pb}$ instead of
$C_{AB}$.)

Unfolded equations of motion of a spin-$s$ massless field are
\cite{Ann}
\be
\label{CON1}
R_{\ga_1\ldots \ga_n,\pa_1\ldots\pa_m} =\eta
\delta^0_n \overline{H}^{\pa_{2s-1}\pa_{2s}}
\overline{C}_{\pa_1\ldots\pa_{2s}}
+\bar  \eta \delta_m^0 {H}^{\ga_{2s-1}\ga_{2s}}{C}_{\ga_1\ldots\ga_{2s}}\,,\qquad
n+m=2(s-1)
\ee
and
\be
\label{CON2}
D^{tw}C_{\ga_1\ldots \ga_n,\pa_1\ldots\pa_m}=0\,,\qquad n-m = 2s\,,\qquad
D^{tw}
\overline{C}_{\ga_1\ldots \ga_n,\pa_1\ldots\pa_m}=0\,,\qquad m-n = 2s\,.
\ee
Here the HS field strength and twisted adjoint covariant
derivative have the form
\be\label{R1}
R_{\ga_1\ldots \ga_n,\pa_1\ldots\pa_m}:=
D^L\omega_{\ga_1\ldots \ga_n,\pa_1\ldots\pa_m}
+n\lambda e_{\ga_1}{}^{\pa_{m+1}}\wedge
\omega_{\ga_2\ldots \ga_n,\pa_1\ldots\pa_{m+1}}
+m \lambda e^{\ga_n+1}{}_{\pa_1}\wedge
\omega_{\ga_1\ldots \ga_{n+1},\pa_2\ldots\pa_{m}}\,,
\ee
\be
\label{bc2}
D^{tw}C_{\ga_1\ldots\ga_n\,,\pa_1\ldots \pa_m} :=
 D^L{C}_{\ga_1\ldots\ga_n\,,\pa_1\ldots \pa_m}
-\f{i}{2}\lambda
(e^{\gamma\dot{\delta}}{C}_{\ga_1\ldots\ga_n\gamma\,,
\pa_1\ldots \pa_m\dot{\delta}}
-n m  e_{\alpha_1\dot{\alpha}_1} {C}_{\alpha_2
\ldots \ga_n,\,\dot{\alpha}_2 \ldots \pa_m}),
\ee
where the indices $\ga$ and $\pa$ are (separately) symmetrized and
Lorentz covariant derivative $D^L$ is
\be
\label{lc}
D^L \psi_\ga := d\psi_\ga + \go_\ga{}^\gb \psi_\gb\q
D^L \overline{\psi}_\pa := d\overline{\psi}_\pa + \overline{\go}_\pa{}^\pb
\overline{\psi}_\pb\,.
\ee
$H^{\ga\gb}=H^{\gb\ga}$ and $\overline{H}^{\pa\pb} =
\overline{H}^{\pb\pa}$ are the basis 2-forms
\be
\label{H}
H^{\ga\gb} := e^{\ga}{}_\pa \wedge e^\gb{}^\pa\,,\qquad
\overline {H}^{\pa\pb} := e_{\ga}{}^\pa\wedge e^{\ga\pb}\,.
\ee
The phase parameters $\eta$ and $\bar{\eta}$ ($\eta \bar{\eta}=1$)
are introduced for the future convenience. Although at the linearized
level they can be absorbed into redefinition of mutually conjugated
$C$ and $\bar C$, they become nontrivial beyond the linearized
approximation \cite{Ann} where the cases of $\eta=1$ and $\eta=i$
correspond to two parity symmetric types of HS theories often referred to as,
 respectively, $A$ and $B$ model\cite{Sezgin:2003pt}.

Formulae are simplified  in terms of  generating functions
\bee
\label{gg}
A (y,\bar{y}\mid x)
=i\sum_{n,m=0}^{\infty}
\frac{1}{n!m!}
{y}_{\alpha_1}\ldots {y}_{\alpha_n}{\bar{y}}_{{\pb}_1}\ldots
{\bar{y}}_{{\pb}_m
} A{}^{\alpha_1\ldots\alpha_n}{}_,{}^{{\pb}_1
\ldots{\pb}_m}(x)\,
\eee
with $A=\omega, C, \overline{C}, R$ etc. In particular,
\be
\label{RRR}
R (y,\bar{y}| x) =D^{ad}\omega(y,{\bar{y}}|x) =
D^L \omega (y,\bar{y}| x) -
\lambda e^{\ga\pb}\Big (y_\ga \frac{\partial}{\partial \bar{y}^\pb}
+ \frac{\partial}{\partial {y}^\ga}\bar{y}_\pb\Big )
\omega (y,\bar{y} | x) \,,
\ee
\be
\label{tw}
D^{tw} C(y,{\bar{y}}|x) =
D^L C (y,{\bar{y}}|x) +\f{i}{2}\lambda e^{\ga\pb}
\Big (y_\ga \bar{y}_\pb -\frac{\partial^2}{\partial y^\ga
\partial \bar{y}^\pb}\Big ) C (y,{\bar{y}}|x)\,,
\ee
\be
\label{dlor}
D^L A (y,{\bar{y}}|x) =
d A (y,{\bar{y}}|x) -
\Big (\go^{\ga\gb}y_\ga \frac{\partial}{\partial {y}^\gb} +
\overline{\go}^{\pa\pb}\bar{y}_\pa \frac{\partial}{\partial \bar{y}^\pb} \Big )
A (y,{\bar{y}}|x)\,.
\ee

As a consequence of the $AdS_4$ zero-curvature equation (\ref{ads}),
the covariant derivatives $D^{ad}$ and $D^{tw}$ are flat, \ie
\be\nn
(D^{ad})^2=(D^{tw})^2 =0\,.
\ee
These conditions imply consistency of equations (\ref{CON1}) and (\ref{CON2})
(\ie compatibility with $d^2=0$) and
gauge invariance of the field strength (\ref{RRR})
(and hence free  equations (\ref{CON1}))
under Abelian HS gauge transformations
\be
\label{hsgau}
\delta \omega(y,\bar{y}|x) = D^{ad} \epsilon(y,\bar{y}|x)\,.
\ee
It is important that consistency of the equations is not
spoiled by the $C$-dependent terms in (\ref{CON1}). As explained
in \cite{333}, this means that the latter correspond to
Chevalley-Eilenberg cohomology of $sp(4,{\mathbb R})$ with
coefficients in the corresponding infinite-dimensional module.

In Eqs.~(\ref{CON1}), (\ref{CON2}), a spin $s$ field
is described by the set of gauge 1-forms
$\omega{}^{\alpha_1\ldots\alpha_n}{}_,{}^{{\pb}_1
\ldots{\pb}_m}(x)$ with $n+m=2(s-1)$ (for $s\geq 1$)
and  0-forms $C{}^{\alpha_1\ldots\alpha_n}{}_,{}^{{\pb}_1
\ldots{\pb}_m}(x)$ with $n-m =2s$ along with their conjugates
$\overline{C}{}^{\alpha_1\ldots\alpha_n}{}_,{}^{{\pb}_1
\ldots{\pb}_m}(x)$ with $m-n =2s$. Indeed, it is easy to see
that the field equations (\ref{CON1}) and (\ref{CON2})
for such a set of fields  form a subsystem for any $s$.

For example, a spin zero field is described by a
set of multispinors
${C}{}^{\alpha_1\ldots\alpha_n}{}_,{}^{{\pb}_1 \ldots{\pb}_n}(x)$
which is equivalent to the set of $4d$ symmetric tensors
$C_{m_1\ldots m_n}(x)$ satisfying the tracelessness condition
(\ref{trace}). This property extends to all spins in the sense that
all multispinors $C{}^{\alpha_1\ldots\alpha_n}{}_,{}^{{\pb}_1
\ldots{\pb}_m}(x)$ describe traceless Lorentz tensors. This
follows from the Penrose formula that any
$p_{\ga\pa}=p_\ga \bar p_\pa$ is null \cite{penr}.
Indeed, unfolded equations
just express $p_{\ga\pa}\sim \f{\p}{\p x^{\ga\pa}}$ via
$p_\ga \bar p_\pa$ with
$p_\ga \sim \f{\p}{\p y^{\ga}}$, $\bar p_\pa \sim
 \f{\p}{\p \bar y^{\pa}}$ (modulo mass-like terms proportional to
$\lambda^2 y^{\ga} y^\pa$ necessary in $AdS$ background).

Dynamical massless fields are
\begin{itemize}
\item
$C(x)$ and $\overline{C}(x)$ for two spin 0 fields,
\item
$C_\ga(x)$ and $\overline{C}_\pa (x)$ for a massless spin 1/2 field,
\item
$\omega_{\ga_1\ldots\ga_{s-1},\pa_1\ldots\pa_{s-1}}(x)$
for an integer spin $s\geq 1 $ massless field,
\item
 $\omega_{\ga_1\ldots\ga_{s-3/2},\pa_1\ldots\pa_{s-1/2}}(x)$
and its complex conjugate
 $\omega_{\ga_1\ldots\ga_{s-1/2},\pa_1\ldots\pa_{s-3/2}}(x)$
for a half-integer spin $s\geq 3/2$ massless field.
\end{itemize}
All other fields  are auxiliary, being
expressed via derivatives of the dynamical massless fields
by  Eqs.~(\ref{CON1}), (\ref{CON2}).

Pattern of the unfolded  massless field equations
is expressed by the so called  Central On-Shell Theorem \cite{Ann}
stating that  Eqs.~(\ref{CON1}), (\ref{CON2})
 express all auxiliary fields in terms of derivatives of the dynamical
fields, imposing on the latter massless field equations  equivalent to those
of Fronsdal \cite{Frhs} and Fang and Fronsdal \cite{Frfhs}.

More in detail, the meaning of Eqs.~(\ref{CON1}), (\ref{CON2})
is as follows. Eqs.~(\ref{CON2}) are
independent from (\ref{CON1})
for spins $s=0$ and $s=\half$ and partially independent
for $s=1$ but become consequences of (\ref{CON1}) for  $s>1$.
Eqs.~(\ref{CON1}) express the holomorphic
and antiholomorphic components of spin $s\geq 1$ 0-forms
$C(y,\bar{y}|x)$ via derivatives of the
gauge 1-forms  $\go(y,\bar{y}|x)$. This identifies
the spin $s\geq 1$  holomorphic
and antiholomorphic components of the 0-forms
$C(y,\bar{y}|x)$ with the Maxwell tensor,
on-shell Rarita-Schwinger curvature, Weyl tensor and their HS
counterparts considered already in the seminal works
\cite{Weinberg:1965rz,penr}. In addition,  Eqs.~(\ref{CON1}) impose
usual (second-order for bosons and first-order for fermions)
 field equations on the spin $s>1$ massless gauge fields
so that Eqs.~(\ref{CON2}) become their
consequences by virtue of Bianchi identities. Dynamical equations
for spins $s\leq 1$ are contained in Eqs.~(\ref{CON2}).

Although the system (\ref{CON1}), (\ref{CON2})
is consistent at the free field level, its nonlinear
 extension requires  doubling of fields
 \cite{Ann,con,more}.
This is achieved by introducing the fields
\be\nn
\omega^{ii} (y,\bar{y}| x)\,,\qquad
C^{i 1-i} (y,\bar{y}| x)\,,\qquad i=0,1
\ee
such that $\omega^{ii} (y,\bar{y}| x)$ are selfconjugated,
while $C^{01} (y,\bar{y}| x)$ and
$C^{10} (y,\bar{y}\mid x)$ are conjugated to one another,
\be
\nn
\overline{\omega^{ii} (y,\bar{y}| x)}=\omega^{ii} (\bar{y},{y}| x)\q
\overline{C^{i\,1-i} (y,\bar{y}| x)} = C^{1-i\,i} (\bar{y},{y}| x)\,.
\ee
The unfolded system for the doubled set of fields is
 \be
\label{CON12}
R^{ii}(y,\overline{y}\mid x) =  \eta\overline{H}^{\pa\pb}
\f{\p^2}{\p \overline{y}^{\pa} \p \overline{y}^{\pb}}
{C^{1-i\,i}}(0,\overline{y}\mid x) + \bar  \eta H^{\ga\gb}
\f{\p^2}{\p {y}^{\ga} \p {y}^{\gb}}
{C^{i\,1-i}}(y,0\mid x)\,,
\ee
\be
\label{CON22}
D^{tw}C^{i \,1-i}(y,\overline{y}\mid x) =0\,.
\ee
Note that now all components of the expansions (\ref{gg})
of $C^{i\,1-i} (y,\bar{y}\mid x)$ contribute to
Eqs.~(\ref{CON12}), (\ref{CON22}),
 while in Eqs.~(\ref{CON1}), (\ref{CON2})
with the single HS 1-form $\go(y,\by)$
only those components of $C(y,\by)$
($\overline{C}(y,\by)$) contributed, that carried at
least as many $y(\bar y)$ as $\bar y(y)$.

In the standard formulation of the $4d$ nonlinear HS gauge theory
\cite{more,Vasiliev:1999ba},
 the field doubling  results from the dependence on the
Klein operators $k$ and $\bar{k}$ that have the properties
\be
\label{kk}
k w^\ga = -w^\ga k\,,\quad
k \bar w^\pa = \bar w^\pa k\,,\quad
\bar k w^\ga = w^\ga \bar k\,,\quad
\bar k \bar w^\pa = -\bar w^\pa \bar k\,,\quad k^2=\bar k^2 = 1\,,\quad
k\bar k = \bar k k\,,
\ee
where $w^\ga =y^\ga$, $\bar w^\pa = \bar y^\pa$.
All fields are packed into 1-forms
\be\nn
\omega(y,\overline{y};k,\overline{k}\mid x)=\sum_{ij = 0,1}
(k)^i (\overline{k})^j \omega^{ij}(y,\overline{y}\mid x)\,
\ee
and 0-forms
\be\nn
C(y,\overline{y};k,\overline{k}\mid x)=\sum_{ij = 0,1}
(k)^i (\overline{k})^j C^{ij}(y,\overline{y}\mid x)\,.
\ee
Now both adjoint and twisted adjoint covariant derivatives
result from different sectors of the adjoint
covariant derivative in the Weyl algebra extended by the
Klein operators.

Massless fields are those with
\be\nn
\omega(y,\overline{y};-k,-\overline{k}\mid x)=
\omega(y,\overline{y};k,\overline{k}\mid x)\,,\qquad
C(y,\overline{y};-k,-\overline{k}\mid x)=-
C(y,\overline{y};k,\overline{k}\mid x)\,.
\ee
The fields with the opposite oddnesses in the Klein operators
are topological, carrying at most a finite number of
degrees of freedom per an irreducible  subsystem
\cite{aux}.

Truncating out fermions, it is possible to consider
a system with bosonic fields in which every integer spin
appears once.
This is achieved via projection of the theory with the help of projectors
\be
\label{Pi}
\Pi_\pm = \half (1\pm k\bar k)\,,
\ee
which are central in the bosonic HS theory. HS gauge fields
and Weyl 0-forms of the bosonic theory are
$\omega_{\pm} =\half (\omega^{00} \pm \omega^{11})$ and
${C}_\pm =\half ({C}^{01} \pm {C}^{10})$.
Bosonic HS theories can be
further truncated to the minimal  system that only contains
even spins \cite{Vasiliev:1999ba}.

\section{Conserved currents and massless equations}
\label{currents}

As observed in \cite{BLS,BHS,333},
conformal invariant massless equations are naturally formulated
in the spaces $\M_M$  with matrix coordinates $X^{AB}=X^{BA}$
($A,B=1,\ldots M$). More precisely, $3d$ Minkowski space-time
coincides with $\M_2$ while $4d$ Minkowski space-time is a subspace
of the ten-dimensional space $\M_4$.
In the both cases unfolded massless field equations are
\be
\label{r1}
d X^{AB} \Big (\frac{\partial}{\partial  X^{AB}} \pm
\frac{\partial^2}{\partial Y^A \partial Y^B}\Big ) C_\pm(Y|X)=0\,,
\ee
where  $\pm$  is introduced for the future convenience.

In \cite{tens2}, Eq.~(\ref{r1}) was extended to so-called
rank-$r$ unfolded equations
\be
\label{co}
d X^{AB} \Big (\frac{\partial}{\partial  X^{AB}}\pm\eta^{ij}
\frac{\partial^2}{\partial Y^{iA} \partial Y^{jB}}\Big ) C^r_\pm(Y|X)=0\,,
\ee
where $i,j=1,\ldots, r$ and $\eta^{ij}=\eta^{ji}$ is some nondegenerate metric.
Higher-rank systems inherit all symmetries of the underlying lower-rank system
simply because they correspond to the tensor
product of the lower-rank representation.
In particular,  higher-rank systems are conformal once
the underlying lower-rank systems were. In the basis where $\eta^{ij}$
is diagonal, the higher-rank
equation (\ref{co}) is satisfied by the product of rank-one fields
\be
C^r (Y_i|X) = C_1(Y_1|X)C_2(Y_2|X)\ldots C_r(Y_r|X)\,.
\ee

A  rank-$r$ system in $\M_M$ can be further  extended to
the rank-one system (\ref{r1}) in the larger space $\M_{rM}$
with coordinates $X^{AB}_{ij}$ via reinterpretation of the
twistor coordinates
\be
Y^A_i \to Y^{\widetilde A}\q \widetilde A = 1,\ldots, rM\,.
\ee
The diagonal embedding of $\M_M$ into $\M_{rM}$ is
\be
X^{AB}_{11}= X^{AB}_{22}=\ldots = X^{AB}_{rr} =X^{AB}\,.
\ee

In particular, the $M=2$ rank-two system  extends to
the $M=4$ rank-one system.
Group theoretically, this provides a realization of the
Flato and Fronsdal theorem \cite{FF} that relates
tensor products of $3d$ conformal fields (singletons) to
infinite towers of $4d$ massless fields of all spins.
The key fact underlying
$AdS_4/CFT_3$ holographic duality in HS theories
is that, as shown in \cite{tens2}, rank-two systems just
describe conserved currents. Direct identification of
$3d$ conserved conformal currents with $4d$ massless fields
provides an example of holographic duality via unfolded dynamics.

\renewcommand{\W}{W}
\renewcommand{\U}{U}

Let us recall following \cite{gelcur} how rank-two
equations give rise to conserved currents.
The rank-two equation can be rewritten in the form
\be \label{unfol2uyR}\left\{
\,\f{\p}{\p X{}^{AB}}  -  \f{\p^2}{\p Y^{(A}\p U^{B)}}
 \right\} T(U,\,Y| X) =0\,,
 \ee
where $T(U,\,Y| X)$ will be called generalized stress tensor. In particular,
Eq.~(\ref{unfol2uyR}) is obeyed by the bilinear substitution
\be\label{Tbilin}
T(U,\,Y\,|X) =\sum_{i=1}^N C_{+\,i}( Y-\U|X)\,
C_{-\,i}( {U}+Y|X)\,,
\ee
where $C_{\pm\,i}(  Y|X)$ obey (\ref{r1}).
Rank-two fields can be interpreted as
bilocal fields in the twistor space.
Being seemingly similar to the bilocal
space-time formalism of \cite{Das:2003vw,Koch:2010cy,Jevicki:2011ss},
the twistorial bilocal formalism is in many respects more efficient.

Since Eq.~(\ref{unfol2uyR}) has unfolded form, its dynamical
pattern can be analyzed with the help of $\sigma_-$-cohomology
techniques with
\be
\sigma_- = -dX^{AB} \f{\p^2}{\p Y^{A}\p U^{B}}\,.
\ee
The result is  that dynamical currents (primaries), that belong to
$H^0(\sigma_-)$, are \cite{tens2}
\be
\label{JJ}
J(U|X) = T(U,0\,|X)\q \tilde J(Y|X) = T(0,Y|X)\,,
\ee
\be
\label{asym}
J^{asym}(U,Y|X) =  ( U^A Y^B - U^B Y^A) \Big (\f{\p^2}{\p U^A \p Y^B}
T(U,Y|X)\Big |_{U^A=Y^A=0}\Big )\,.
\ee

In the $3d$ case of $M=2$ where $A,B\to \ga,\gb$, $J(U|X)$ generates
$3d$ currents of all integer and half-integer spins \be J(U|X)=
\sum_{2s=0}^\infty U^{\ga_1}\ldots U^{\ga_{2s}} J_{\ga_1\ldots
\ga_{2s}}(X)\,
 \ee and \be \label{Jasym} J^{asym}(U,Y|X) = U_\ga
Y^\ga J^{asym} (X)\,. \ee

Differential equations, which follow from
Eq.~(\ref{unfol2uyR}), are associated with
$H^1(\sigma_-)$ found in \cite{tens2} for general $M$.
For $M=2$, the structure of $H^1(\sigma_-)$
is greatly simplified so that the resulting field equations
amount solely to the conventional conservation condition
\be
\label{ccc}
\f{\p}{\p X^{\ga\gb}}\f{\p^2}{\p U_\ga \p U_\gb} J(U|X) =0\q
\f{\p}{\p X^{\ga\gb}}\f{\p^2}{\p Y_\ga \p Y_\gb} \tilde J(Y|X) =0\,.
\ee

To define conserved charges, it is convenient to Fourier
transform $T(\U,\,Y\,|X)$ to
\bee
\label{Fourier} \widetilde T ( \W,\,Y|X)\,\,=\,\, 
(2\pi)^{-M/2}\int\limits_{\mathbb{R}^M}  
d^{\,M}\U\,\,\, \exp\left(-i\,\,\,
\W_C\,\U{}^C\right)\,T(\U,\,Y\,|X)\,,
 \eee
which satisfies the following {\it current equation}
\be \label{unfol2_Fur} \left( \, \f{\p}{\p X{}^{AB}}  +
i \W_{(A}
 \f{\p}{\p   Y{}{\,}{}^{B)}}
 \right)  \widetilde T( \W,\,Y|X)=0\,.
\ee
 The key fact is that a $2M$--form
\bee\label{varpi}
\Omega(T)\,=\,
 \,\left(d\,\W_A \wedge
\Big(i\,\W{}_B{} d\, X{}^{AB} -    d\,Y{}^A
\Big)\right)^{\,M}\,\, \widetilde T(\W,\,Y\,|X)
\eee
is closed in
$M_M \times
\mathbb{R}^{M}(\W_B)\times  \mathbb{C}^{M}(Y^A)
$
provided that
$\widetilde T(\W,\,Y\,|X)$ obeys (\ref{unfol2_Fur}).

\renewcommand{\Z}{X}
\renewcommand{\gY}{Y}

Indeed, from (\ref{unfol2_Fur}) and (\ref{varpi}) it follows that
\bee\nn &&\left (d\,\W_{A} \f{\p}{\p \W_{A}}+
d\,\Z{}^{AB} \f{\p}{\p \Z{}^{AB}} +  d\,\gY^{A} \f{\p}{\p
\gY^{A}}\right )\wedge
\Omega^{2M}(T(\W,\,\gY\,|\Z ))\,=
\\ \nn
= &&\ls\left ( d\,\W_{A}
\f{\p}{\p \W_{A}} \,- \,\Big(i\,\W{}_B{} d\,\Z{}^{AB} -d\,\gY{}^A
 \Big ) \f{\p}{\p \gY^{A}}\right)\wedge \Omega^{2M}(T(\W,\,\gY\,|\Z))=0\,.
\eee
As a result, the charge
\be
\label{Q} q= q(T)=\int_{\Sigma^{2M}} \Omega^{2M}(T)
\ee
is independent of local variations of a $2M$-dimensional
integration surface $\Sigma^{2M}$ on solutions of
(\ref{unfol2uyR}).
In particular, for functions that
{decrease  fast enough at space infinity}, it is independent of the
time parameter in $\M_M$, hence being conserved.

A remarkable output of this construction  \cite{gelcur}
is that it makes it possible
to  express conserved charges as integrals over the twistor
space $\mathbf{T}$ at any point of space-time.

Since  (\ref{unfol2_Fur}) is a first-order linear PDE system,
its  solutions form a commutative
algebra $\R$, \ie a linear combination of products of any regular
solutions of (\ref{unfol2_Fur}) is also a solution.  $\R$
is the algebra of functions $\eta$ of the form
\be \label{par} \eta(\W,\,\gY\,|\Z) =
\varepsilon(\W_A,\,\gY^C \,-i\,\Z{}^{CB}\,\W_B)\,
\ee
with regular  $\varepsilon(\W,\gY)$.
As a result, Eq.~(\ref{varpi}) generates conserved
currents for $\widetilde T_\eta(\W,\,\gY\,|\Z) $ of the form
\be\label{eta_f}
\widetilde T_\eta(\W,\,\gY\,|\Z) = \eta(\W,\,\gY\,|\Z)
\widetilde T(\W,\,\gY\,|\Z)\,,
 \ee
 where
 $\eta(\W,\,\gY\,|\Z)$  (\ref{par}) is a polynomial
 representing a parameter of global HS symmetry.
 The charges $q(\widetilde T_\eta)$ with various
$\eta(\W,\,\gY\,|\Z)$  generate
the full set of conformal HS conserved charges. In particular,
at $M=2$, formula (\ref{Tbilin}) generates all conserved
charges for free $3d$ massless fields.

\section{Nonlinear HS equations in $AdS_4$}
\label{Nonlinear Higher-Spin Equations}

In this section,  we first recall standard formulation
of nonlinear $4d$ HS  equations of \cite{more} and then
extend it to a larger space with spinor coordinates. In the sequel,
wedge products are implicit.

\subsection{Standard formulation}
\label{Standard formulation}

The key element of the construction of \cite{more} is the doubling of auxiliary
Majorana spinor variables $Y_A$ in the HS 1-forms and 0-forms
\be
\go(Y;\ck|x)\longrightarrow W(Z;Y;\ck|x)\,,\qquad
C(Y;\ck|x)\longrightarrow B(Z;Y;\ck|x)
\ee
supplemented with equations which determine  dependence on the
additional variables $Z_A$ in terms of ``initial data"
\be
\label{inda}
\go(Y;\ck|x)=W(0;Y;\ck|x)\,,\qquad C(Y;\ck|x)= B(0;Y;\ck|x).
\ee
To this end, we introduce a
spinor field $S_A (Z;Y;\ck|x)$ that carries only pure gauge
degrees of freedom and plays a role of connection with respect to
additional $Z^A$ directions. It is convenient to
introduce anticommuting $Z-$differentials $dZ^A dZ^B=-dZ^B
dZ^A$ to interpret $S_A (Z;Y;\ck|x)$ as a $Z$--1-form,
\be
S=dZ^A S_A (Z;Y;\ck|x) \,.
\ee
The variables $\ck=(k,\bar{k})$ are Klein operators
that satisfy (\ref{kk}) with
 $w^\ga= (y^\ga, z^\ga, dz^\ga )$, $\bar w^\pa =
(\bar y^\pa, \bar z^\pa, d\bar z^\pa )$.

The nonlinear HS equations are \cite{more}
\be
\label{dW}
dW+W*W=0\,,\qquad
\ee
\be
\label{dB}
dB+W*B-B*W=0\,,\qquad
\ee
\be
\label{dS}
dS+W*S-S*W=0\,,
\ee
\be
\label{SB}
S*B=B*S\,,
\ee
\be
\label{SS}
S*S= -i (dZ^A dZ_A + dz^\ga dz_\ga  F_*(B) k\ups +
d\bar z^\dga d\bar
z_\dga \bar F_*(B) \bar k \bu)
\,,
\ee
where
$F_*(B) $ is some star-product function of the field $B$.

The simplest case of linear functions
\be
\label{etaB}
F_*(B)=\eta B \q \bar F_* (B) = \bar\eta B\,,
\ee
where $\eta$ is some phase factor (its absolute value can be absorbed into
redefinition of $B$) leads to a class of pairwise nonequivalent nonlinear HS
theories. The particular cases
of $\eta=1$ and $\eta =\exp{\f{i\pi}{2}}$ are especially interesting, corresponding
to so called $A$ and $B$ HS models. These two cases are distinguished
by the property that they  respect parity \cite{Sezgin:2003pt}.

The associative star product $*$ acts on functions of two
spinor variables
\be
\label{star2}
(f*g)(Z;Y)=\frac{1}{(2\pi)^{4}}
\int d^{4} U\,d^{4} V \exp{[iU^A V^B C_{AB}]}\, f(Z+U;Y+U)
g(Z-V;Y+V) \,,
\ee
where
$C_{AB}=(\epsilon_{\ga\gb}, \bar \epsilon_{\dga\dgb})$
is the $4d$ charge conjugation matrix and
$ U^A $, $ V^B $ are real integration variables. It is
normalized so that 1 is a unit element of the star-product
algebra, \ie $f*1 = 1*f =f\,.$ Star product
(\ref{star2}) provides a particular
realization of the Weyl algebra
\be
[Y_A,Y_B]_*=-[Z_A,Z_B ]_*=2iC_{AB},,\qquad
[Y_A,Z_B]_*=0
\ee
($[a,b]_*=a*b-b*a$) resulting from the normal ordering
 with respect to the elements
\be
\label{oscrel}
   b_A = \frac{1}{2i} (Y_A - Z_A )\,,\qquad
   a_A = \frac{1}{2} (Y_A + Z_A )\,,\qquad
\ee
which satisfy
\be
\label{com a}
   [a_A, a_B]_*=[b_A, b_B]_* =0 \,,\quad
   [a_A, b_B]_* =C_{AB} \,
\ee
and can be interpreted as creation and annihilation operators as
is most evident from the relations
\be
\label{pr a}
   b_A * f(b, a) =
      b_A f(b, a) \,,\qquad
   f(b, a) * a_A =
      f(b, a)  a_A \,.
\ee

{}From (\ref{star2}) it follows that functions $f(Y)$  form a proper subalgebra which is the centralizer of the elements
$Z_A$. For $Z$-independent $f(Y)$ the star product (\ref{star2})
takes the form of the Weyl star product.

An important property of the star product (\ref{star2}) is that it
admits the inner Klein operator
\be
\Upsilon = \exp i Z_A Y^A \,,
\ee
which behaves as $(-1)^{N} ,$ where $N$ is the spinor number
operator. One can easily see that
\renewcommand{\U}{\Upsilon}
\be
\U *\U =1,
\ee
\be
\label{[UF]}
\U *f(Z;Y)=f(-Z;-Y)*\U\,,\quad
\ee
and
\be
\label{Uf}
(\U *f)(Z;Y)=\exp{i Z_A Y^A }\, f(Y;Z) \,.
\ee
The left and right inner Klein operators
\be
\label{kk4}
\ups =\exp i z_\ga y^\ga\,,\qquad
\bu =\exp i \bar{z}_\dga \bar{y}^\dga\,,
\ee
 which enter Eq.~(\ref{SS}), act analogously
on  undotted and dotted spinors, respectively
\be
\label{uf}
\!(\ups *f)(z,\!\bar{z};y,\!\bar{y})\!=\!\exp{i z_\ga y^\ga }\,\!
f(y,\!\bar{z};z,\!\bar{y}) ,\quad\! (\bu
*f)(z,\!\bar{z};y,\!\bar{y})\!=\!\exp{i \bar{z}_\dga \bar{y}^\dga
}\,\! f(z,\!\bar{y};y,\!\bar{z}) ,
\ee
\be
\label{[uf]}
\ups *f(z,\bar{z};y,\bar{y})=f(-z,\bar{z};-y,\bar{y})*\ups\,,\quad
\bu *f(z,\bar{z};y,\bar{y})=f(z,-\bar{z};y,-\bar{y})*\bu\,,
\ee
\be
\ups *\ups =\bu *\bu =1\q \ups *\bu = \bu*\ups\,.
\ee

To analyze Eqs.~(\ref{dW})-(\ref{SS}) perturbatively,
one has to linearize them around some vacuum solution.
The simplest choice is
\be
W_0(Z;Y|x)= W_0(Y|x)\q S_0(Z;Y|x) = dZ^A Z_A\q B_0=0\,,
\ee
where $W_0(Y|x)$ is some solution of the flatness condition
\be
dW_0(Y|x) + W_0(Y|x)*W_0(Y|x)=0\,.
\ee
 $W_0(Y|x)$ bilinear in $Y^A$ describes $AdS_4$. Linearization of the system
(\ref{dW})-(\ref{SS}) around this vacuum  just
reproduces the free  field equations (\ref{CON1}), (\ref{CON2})
(for more detail see \cite{more,Vasiliev:1999ba}).

In the purely bosonic case where all fermion fields are zero,
the operator $k\bar k$ remains central in the full nonlinear
system. As a result, bosonic sector of the system
(\ref{dW})-(\ref{SS}) decomposes into two independent subsectors
singled out by the projectors $\Pi_\pm$ (\ref{Pi}).

\subsection{Spinor coordinates}
\label{ferext}

An important feature of the system (\ref{dW})-(\ref{SS})
is that Eqs.~(\ref{dW})-(\ref{dS}), that contain space-time
differential $d$, are flatness conditions. As a result, the flows
along space-time coordinates commute to Eqs.~(\ref{SB}),
(\ref{SS}). This has two  consequences. One is that  nontrivial
dynamics is hidden entirely in the noncommutative twistor space of
$Z$ and $Y$ \cite{alg}. Another is that the system remains
consistent if  original space-time coordinates $x^{\ga\pa}$
are extended to a larger space. Of course, if
the differential is extended as in (\ref{dext}) with the same vacuum
connection, additional equations
will simply mean that, up to gauge ambiguity, all fields
are independent of the coordinates $\hat X$. However, the situation
becomes more interesting if pullback of a vacuum connection
to additional directions is nonzero.

As explained in Section \ref{vacuum},
this can be achieved by introducing a connection with respect to
some symmetry algebra $h$ in the system. So far, vacuum
connection was introduced  for the $AdS_4$ algebra
$sp(4|\mathbb{R})$. We believe that in the context of holographic
interpretation of the $AdS_4$ HS theory it may be useful to extend
$sp(4|\mathbb{R})$ to the Lie algebra with
the generators
\be
\label{hei}
T_{AB}= -\f{i}{2} Y_A Y_B\q t_A=  Y_A\,
\ee
obeying commutation relations
\be
[T_{AB}\,,T_{CD} ] = C_{BC} T_{AD} +C_{AC} T_{BD}
+C_{BD} T_{AC} +C_{AD} T_{BC}\,,
\ee
\be
\label{hcom}
[T_{AB}\,,t_C ] = C_{BC} t_A +C_{AC} t_B\q
[t_A\,,t_B] = 2i C_{AB}\,.
\ee
(The central element on the r.h.s. of the second relation
(\ref{hcom}), which can be identified with $\hbar$ in the
Heisenberg algebra $h_4$ spanned by $T_A$, is set to unity.)
Following  \cite{gvtw}, we  call this Lie algebra
$sph(4|\mathbb{R})$. It  should not be
confused with the superalgebra $osp(1,4)$ where $Y_A$ are
treated as supergenerators.
 Clearly, $sph(4|\mathbb{R})=
sp(4|\mathbb{R})\hsum h_4$. Note that $sph(4|\mathbb{R})$ is
a parabolic subalgebra of $sp(6|\mathbb{R})$.

This generalization is aimed at the extension to the action level
of the construction of \cite{gelcur} explained in Section
\ref{currents} where  conserved currents
were represented by closed forms in the correspondence space
unifying space-time and spinor coordinates. As shown in \cite{gvtw},
relevant geometry naturally results from the formulation of HS theory
in an appropriate coset space of the group $SpH(4|\mathbb{R})$.
The idea is to
introduce additional commutative coordinates $u^{\underline A}$ associated with
additional generators in $sph(4|\mathbb{R})$ compared to
$sp(4|\mathbb{R})$. Namely, we set
\be
X=(x^{\underline{\ga}\underline{\pa}}, u^{\underline A})\q u^{\underline A}=(u^{\underline \ga}
\,,\bar u^{\underline \pa} )\,.
\ee
Correspondingly,  the space-time HS connection $W_x(Z;Y;\K|x)=
dx^\un W_\un (Z;Y;\K|x)$ is extended to
\be
W_X(Z;Y;\K|X) = W_x(Z;Y;\K|X)+ W_u(Z;Y;\K|X)\q
\ee
\be
W_x(Z;Y;\K|X)=
dx^\un W_\un (Z;Y;\K|X)\q W_u(Z;Y;\K|x)=
du^{\underline A} W_\uA (Z;Y;\K|X)\,.
\ee

A vacuum connection can be chosen in the form
\be
W_{0x}(Y|x) = \f{i}{4}
W^{AB}_0(x)Y_A Y_B\q W_{0u}= du^\uA W_{0\uA}{}^A(x) Y_A+
idu^\uA u^\uB C_{\uA\uB}\,,
\ee
where $W_\uA{}^A(x)$ is a set of Killing spinors enumerated by
the label $\uA$, that satisfy the covariant constancy
condition
\be
dW_{0\uA}{}^A(x) -W_0^{A}{}_B(x)  W_{0\uA}{}^B(x) =0\,.
\ee
As a consequence,
\be
C_{\uA \uB} = W_{0\uA}{}^A(x) W_{0\uB}{}^B(x) C_{AB}
\ee
is some constant antisymmetric matrix. Requiring
\be
W_{0\uA}{}^A(x_0) = \delta_\uA^A
\ee
at some point $x_0$, we achieve that $C_{\uA\uB}(x)=C_{AB}$ for
any $x$.

As explained in Section \ref{actions}, a HS action should be
described by some $Q$--closed $4$--form where $Q$ is the operator
(\ref{qdif}) associated with the unfolded form of HS equations resulting from the
perturbative analysis of Eqs.~(\ref{dW})-(\ref{SS}). In other words, the
action should be $d$--closed
by virtue of these equations. Since Eqs.~(\ref{dW})-(\ref{SS}) are
insensitive to particular detail of space-time, this should be
true for any space-time coordinates that can be
introduced within unfolded formulation.
The idea of the spinor extension is to use coordinates $u^\uA$
instead of $x^\un$ in the computations involving
bulk-to-boundary propagators, which are expected to be
simpler in the spinor space than in $AdS_4$.
A toy model for  this mechanism is provided by the
evaluation of conserved charges along the lines of \cite{gelcur}.

\section{$AdS_4$  HS theory as $3d$ conformal HS theory}
\label{ads4cft3}

As discussed in Section \ref{Duality}, unfolded formulation
allows one to choose freely one or another space-time interpretation
of the theory. To see HS $AdS_4/CFT_3$ holographic duality,
consider  pullback of all
space-time differential forms (\ie curvatures and connections)
to some $3d$ surface $\Sigma  \in AdS_4$. This gives a subsystem
of the original unfolded system in $AdS_4$ that now can be
interpreted as a $3d$  system on $\Sigma$, being
by construction equivalent to the original
 $AdS_4$ system. In the HS model of interest,
0-forms associated with $AdS_4$ massless fields
 acquire the meaning of $3d$ conserved currents. Indeed, from the $3d$ perspective,
dotted and undotted indices carry equivalent Lorentz representation.
Hence, the $3d$ pullback of Eq.~(\ref{CON22})  gives  Eq.~(\ref{unfol2_Fur}) for
the generalized $3d$ conformal stress tensors. What is not guaranteed
however is that conformal properties are manifest for
dynamical variables inherited from the  $AdS_4$ HS theory.
Let us consider this point in some more detail.

For manifest conformal invariance, it is most convenient to introduce oscillators
\be
y^+_\ga = \half (y_\ga - i \bar y_\ga)\q
y^-_\ga =\half (\bar y_\ga -i y_\ga)
\ee
that satisfy
\be
\label{concom}
[ y^-_\ga\,, y^{+\gb }]_* = \delta_{\ga}^{\gb}\,.
\ee
In the  $AdS_4$ setup, the reality conditions
imply $(y^\pm_\ga)^\dagger = iy^\pm_\ga $. In the conformal setup
the appropriate reality conditions are
\be
(y^-_\ga)^\dagger = y^{+\ga}\,.
\ee

Weyl star-product realization of the $3d$ conformal  algebra $sp(4;\mathbb{R})\sim
o(3,2)$ is
\be
\label{LD}
L^\ga{}_\gb = y^{+\ga}y^-_\gb-\half \delta^\ga_\gb
y^{+\gga}y^-_\gga \q D = \half y^{+\ga}y^-_\ga\,,
\ee
\be
\label{PK}
P_{\ga\gb} =i y^-_\ga y^-_\gb\q
K^{\ga\gb} =-i y^{+\ga} y^{+\gb}\,.
\ee
Here generators of $3d$ Lorentz transformations $L^\ga{}_\gb$ form
$sp(2;\mathbb{R})$. $D$ is the dilatation generator. $P_{\ga\gb}$
and $K^{\ga\gb}$ are generators of translations and special conformal
transformations, respectively.  By (\ref{concom}),
conformal dimension of the HS gauge fields  counts
the difference of the numbers of pluses and minuses
\be
\label{add}
[D,\go(y^\pm|X) ] = \half \left (y^{+\ga}\f{\p}{\p y^{+\ga}} -
y^-_{\ga} \f{\p}{\p y^-_{\ga}} \right )\go (y^\pm|X)\,.
\ee

$3d$ conformal HS algebra coincides with the
$AdS_4$ HS algebra rewritten in terms of oscillators
$y^\pm_\ga$. In this form it was introduced in \cite{Fradkin:1989xt}.
Hence, the pullback $\hat \go(y^\pm|x)$ of the $AdS_4$
HS gauge fields $\go(y^\pm|x)$ to $\Sigma$ just gives the full set
of $3d$ conformal HS gauge fields.

To make contact with the standard approach it is convenient to
foliate $AdS_4$ so that
\be
x^\un=({\bf x}^{\underline{a}},z)\,,
\ee
where $\xx^{\underline{a}}$ are coordinates of leafs
(${\underline{a}}=0,1,2$) while $z$
is a foliation parameter. Let
$\hat W(y^\pm|\xx,z)=d\xx^{\underline{a}}
\hat W_{\underline{a}}(y^\pm|\xx,z)$ be
pullback of $W(y^\pm|x)$ to a leaf at some  $z$.
At every $z$, the original $AdS_4$ HS theory  gives rise
to a $3d$ conformal HS theory with $3d$ conformal HS connections
$\hat W(y^\pm|\xx,z)$.
Similarly,
$\hat W_0(y^\pm|\xx,z)$ inherited from some $AdS_4$ vacuum
connection $ W_0(y^\pm|\xx,z)$ provides a flat connection
of the $3d$ conformal algebra $ sp(4)$ on every leaf.

A less trivial part of the $3d$ reduction
is due to the gluing terms in (\ref{CON12})
and field equations (\ref{CON22}) on the 0-forms $C$.
First of all we observe that Eqs.~(\ref{bc2}), (\ref{tw})
give the $AdS$ deformation of Eq.~(\ref{unfol2uyR}).
Hence, in agreement with $AdS/CFT$ correspondence prescription,
0-forms $C$  describing massless fields in $AdS_4$ should be
interpreted as conserved currents in the $3d$ conformal setup.
However, this correspondence is not quite direct
because the original 0-forms $C$ do not transform properly
under conformal transformations. Indeed, the  dilatation
generator $D$ in the twisted adjoint representation
is realized by anticommutator which gives
a second-order differential operator
\be
\{D\,, C\}_* =\left (
 y^{+\ga}y^-_\ga -\f{1}{4} \f{\p^2}{\p y^{+\ga}\p y^-_\ga}\right ) C
\ee
rather than the
first-order operator (\ref{add}) in the  adjoint
representation. This means that the set of fields $C$ inherited from
the $AdS_4$ theory is not manifestly conformal.

To solve this problem one should change variables from
$C(y^\pm |x)$ to $T(y^\pm |x)$ to achieve that $T(y^\pm |x)$ transforms
properly under dilatations. In fact, this issue
is not specific to the conformal description, having its direct
analogue on the $AdS$ side where
the fields $C(y^\pm |x)$ do not exhibit manifest decomposition
in terms of eigenfunctions of the energy operator $E$ which
is holographically dual to $D$. As shown in \cite{BHS},
the transition from $C(y^\pm |x)$ to the
 basis which diagonalizes  energy is nonlocal, forming
a kind of non-unitary Bogolyubov transform. Similar transformation
in the conformal setup is achieved by the transition from Weyl
to Wick star product with respect to $y^-$ and $y^+$.

Let $f(y^\pm)$ be an element of the Weyl star-product algebra.
The map from the Weyl star product
\be
(f*g) (y^\pm) =\f{1}{\pi^4}\int d^4 u^\pm d^4 v^\pm \exp 2(v^-_\ga u^{+\ga} -u^-_\ga v^{+\ga})
f(y^\pm +u^\pm)g(y^\pm +v^\pm)
\ee
to the Wick star product
\be
\label{Wick}
(f_N\star g_N) (y^\pm) =\f{1}{(2\pi)^2}\int d^4 u^\pm\exp
(-u^-_\ga u^{+\ga})
f_N(y^+, y^- +u^-)g_N(y^+ +u^+, y^-)\,
\ee
is
\be
\label{N}
f_N(y^\pm )= \f{1}{\pi^2}\int d^4 u^\pm
 \exp(-2  u^-_\ga u^{+\ga}) f(y^\pm +u^\pm)\,
\ee
or, equivalently,
\be
\label{Nd}
f_N(y^\pm) = \exp\left (-\half \epsilon^{\ga\gb}
\f{\p^2}{\p y^{-\ga}\p y^{+\gb}}\right ) f(y^\pm)\,.
\ee
Wick star product  has the properties
\be
f_N(y^+)\star g_N(y^\pm) = f_N(y^+) g_N(y^\pm)\q
f_N(y^\pm)\star g_N(y^-) = f_N(y^\pm) g_N(y^-)\,,
\ee
\be
y^-_\ga\star =  y^-_\ga+\f{\p}{\p y^{+\ga}} \q
\star y^+_\ga =  y^+_\ga-\f{\overleftarrow{\p}}{\p y^{-\ga}}\,.
\ee

Let us now apply these formulae to the dilatation operator $D$
(\ref{LD}). First of all, we obtain that
\be
D_N =\half( y^-_\ga y^{+\ga} +1)\,.
\ee
Hence
\be
D_N \star =\half \left ( y^{+\ga} \f{\p}{\p y^{+\ga}}+y^-_\ga y^{+\ga} +1
\right )\q
\star D_N =\half
\left( y^-_{\ga} \f{\p}{\p y^-_{\ga}}+y^-_\ga y^{+\ga} +1\right )\,.
\ee
In the twisted adjoint representation, the
action of $D_N$ therefore is
\be
\{D_N\,,\ldots\}_\star = \half \left (y^{+\ga} \f{\p}{\p y^{+\ga}}
+y^{-\ga} \f{\p}{\p y^{-\ga}}\right )
+ y^-_\ga y^{+\ga} +1\,.
\ee
It remains to introduce
\be
\label{TN}
T(y^\pm|x) = \exp (- y^-_\ga y^{+\ga} )C_N (y^\pm|x)
\ee
to achieve that, in agreement with the interpretation of
$T(y^\pm|x)$ as a $3d$ conformal current,
\be
\label{DNT}
D_N (T(y^\pm)) =\half \left ( y^{+\ga} \f{\p}{\p y^{+\ga}}
+y^{-\ga} \f{\p}{\p y^{-\ga}} +2\right ) T(y^\pm)\,.
\ee

Let us now look more closely at the action of the
translation generator $P_{\ga\gb}$ (\ref{PK}).
To this end we observe that
\be
\label{pm}
k y^\pm_\ga = \mp i y^\mp_\ga k\q
\bar k y^\pm_\ga = \pm i y^\mp_\ga \bar k\,.
\ee
Hence
\be
P_{\ga\gb} T(y^\pm) k =\f{\p^2}{\p y^{+\ga} \p y^{+\gb}}T(y^\pm) k\q
T(y^\pm) k P_{\ga\gb} =-\f{\p^2}{\p y^{-\ga} \p y^{-\gb}}T(y^\pm) k
\ee
and, using $3d$ Cartesian coordinates along with
\be
\f{\p^2}{\p y^{+\ga} \p y^{+\gb}} +\f{\p^2}{\p y^{-\ga} \p y^{-\gb}}
=4 \f{\p^2}{\p y^\ga \p \bar {y}^\gb}\,,
\ee
for the case where the pullback of the $AdS_4$ connection to $\Sigma$
has $3d$ Cartesian form, the resulting equation on the $3d$ 0-forms
 acquires the form of rank-two equation (\ref{unfol2uyR})
\be
\label{tcon}
D_\xx ^{tw} T(y,\bar y|x) = d_\xx T(y,\bar y|x)
+4d\xx^{\ga\gb} \f{\p^2}{\p y^\ga \p \bar {y}^\gb}
T(y,\bar y|x)=0\,.
\ee

According to  the analysis of $\sigma_-$ cohomology
of \cite{tens2} summarized in Section \ref{currents},
Eq.~(\ref{tcon}) describes two sets of conserved currents
of all spins $s>0$ and two spin zero branches distinguished by
their symmetry under $y\leftrightarrow \bar y$. The
symmetric branch is generated by
$J^{sym}(x)=T(0|x)$ that has conformal dimension
$\Delta = 1$. The antisymmetric branch is generated
by $J^{asym}(x)$ (\ref{Jasym}). In accordance with (\ref{DNT}),
$\Delta (J^{asym}(x))=2$. These are just correct conformal dimensions
for scalar currents associated with the two energy branches of the
$AdS_4$ conformal scalar field.

In this setup, the $AdS_4/CFT_3$ HS duality takes place on
every leaf of the $z$--foliation. However,
boundary and bulk fields are related by the nonlocal
transform from Weyl to Wick star product.
Since unfolded equations
relate space-time derivatives to those over $y^\pm$,
nonlocal map in spinor variables is translated to
the space-time nonlocality of the map of $AdS_4$ fields
to conformal ones. Without using twistor
variables it may be difficult to establish precise
correspondence between the $AdS_4$ HS theory and its dual
on any $\Sigma$. However, as shown in the next section, the
holographic correspondence drastically simplifies if $\Sigma$
is $AdS_4$ infinity, just reproducing the standard
$AdS/CFT$ recipe \cite{Gubser:1998bc,Witten:1998qj} where fields
at the boundary of  $AdS_{d+1}$ are
identified with (sources for) currents in $CFT_d$.

Opposite $-+$ ordering choice leads to equivalent
results with the exchange of $y^+$ and $y^-$. This means that
 $D$ changes its sign while $P_{\ga\gb}$ and $K^{\ga\gb}$
should  be reinterpreted as generators of special conformal
transformations and translations, respectively. With these
redefinitions  consideration remains intact.

\section{Holographic locality at infinity}
\label{cinf}

\subsection{Conformal foliation and Poincar{\'e} coordinates}
\label{conff}
Let $M^d$ be a $d$--dimensional conformally flat space-time
with local coordinates $\xx$ and  $w_{\xx}(\xx)=
w^A_{\xx} T_A$ be some flat $o(d,2)$
connection\footnote{That an $o(d,2)$ connection $w_{\xx}(\xx)$ is flat means that
$M^d$ endowed with the metric resulting from the vielbein
associated with the $P_a$ component of $w_{\xx}(\xx)$
is (locally) conformally flat (see, e.g.,
\cite{333}).}
\be
d_\xx w_{\xx}(\xx) + w_{\xx}(\xx)w_{\xx}(\xx) =0\,.
\ee
 Let $D$ be the dilatation generator
among $T_A$ which induces standard $\mathbb{Z}$ grading
on $o(d,2)$ so that
\be
[D\,,T_A ]= \Delta (T_A) T_A\,,
\ee
where $\Delta(T_A)$ is conformal dimension of $T_A$ which
 takes values $\pm 1$
or $0$. Namely,
\be
T_A = (L_{ab}, D, K_a, P_a)\,,
\ee
where conformal dimensions of generators of Lorentz
transformations $L_{ab}$, dilatations $D$, special conformal transformations
$K_a$ and translations $P_a$ are
\be
\label{D}
\Delta{L}=0\q\Delta(D)=0\q \Delta(K) = 1\q \Delta (P) = -1\,.
\ee
A particular flat connection
which corresponds to Cartesian coordinates in $M^d$ is
\be
\label{Cart}
w_{\xx}(\xx)= d\xx^a P_a\,.
\ee

Let us now introduce an additional coordinate $z$ and differential
$dz$ so that $x=(\xx,z)$ be the local coordinates of $AdS_{d+1}$. A
conformally foliated connection $W(x)$ of $AdS_{d+1}$ can be
introduced as follows. The components of $W(x)$ with
differentials $d\xx$ are
 \be \label{fol} W_{\xx}^A(x) T_A =
z^{\Delta(T_A)} w^A_{\xx}(\xx) T_A\,,
\ee while the only nonzero
$dz$ component of the connection is associated with the dilatation
generator $D$, having the form \be W_{z}(x) =  - z^{-1} dz D \,. \ee
Clearly, so defined connection $W(x)$ is flat in (a local chart of)
$AdS_{d+1}$. Poincar{\'e} coordinates result from this construction
applied to the connection (\ref{Cart}). It should be stressed,
however, that this construction works for any $o(d,2)$ flat
connection $w_{\xx}(\xx)$ in $M^d$.
 In particular, if $w_{\xx}(\xx)$ itself is some $AdS_d$ connection
 with  nonzero components in  $o(d-1,2)\in o(d,2)$, it can itself
 be represented in the form of conformal foliation with
another foliation parameter $z_1$, continuing the process of
dimension reduction.

In spinor notations, local coordinates of $AdS_4$ are
\be
\label{xz}
x^{\ga\dga }=(\xx^{\ga\dga},-\f{i}{2} \epsilon^{\ga\dga} z^{-1})\,,
\ee
where the symmetric part of $4d$  coordinates
$\xx^{\ga\dga}=\xx^{\dga\ga}$ is identified with coordinates of
$\Sigma$ while
$z$ is the radial coordinate of $AdS_4$. The appearance of
$\epsilon^{\ga\dga}=-\epsilon^{\dga\ga}$ in the definition of $z$ breaks $4d$ Lorentz
symmetry $sp(2;\mathbb{C})$ to $3d$ Lorentz symmetry
$sp(2;\mathbb{R})$ that acts on the both types of spinor
indices.

Now we are in a position to analyze dynamical content of
$AdS_4$ HS equations at $z\to 0$. We will work in terms of
 Weyl star product inherited from (\ref{star2}) in the sector
 of $y^\pm$ variables. Using  (\ref{Cart}) and (\ref{PK}), (\ref{gg}),
$AdS_4$ connection can be chosen in the form
\be
\label{w0}
W =\f{i}{ z}d\xx^{\ga\gb} y^-_\ga y^-_\gb - \f{dz}{2z} y^-_\ga
y^{+\ga}\,,
\ee
which is equivalent to
\be
\label{flosc}
W =\f{1}{4 z}\left (d\xx^{\ga\gb}(y_\ga y_\gb -
\bar y_\ga \bar y_\gb)
+2i d\xx^{\ga\gb} y_\ga \bar y_{\gb}
+dz y_\ga
\bar y^{\ga}\right )\,.
\ee
 By Eq.~(\ref{gg}),
$AdS_4$ vierbein and Lorentz connection are
\be\label{poframe}
e^{\ga\dga} = \f{1}{2z} dx^{\ga\dga}\q \go^{\ga\gb}=-
\f{i}{4z} d\xx^{\ga\gb}\q \bar \go^{\dga\dgb} =
\f{i}{4z} d\xx^{\dga\dgb}\,.
\ee

\subsection{0-forms}
The unfolded equation on Weyl tensors in $AdS_4$,
which is the covariant constancy condition (\ref{CON22})
in the twisted adjoint representation, decomposes into two
equations with respect to the $3d$ coordinates
$\xx^{\ga\gb}$ and $z$, respectively,
\be
\label{dx}
\Big [d_\xx +\f{i}{z}d\xx^{\ga\gb}\left (y_\ga \f{\p}{\p y^\gb} -
\bar y_\ga \f{\p}{\p \bar y^\gb} + y_\ga\bar y_\gb
- \f{\p^2}{\p y^\ga \p \bar y^\gb}\right )\Big ] C(y,\bar y|\xx,z)=0\,,
\ee
\be
\label{dz}
\Big [d_z +\f{dz}{2z}\left ( y_\ga\bar y^\ga
-\epsilon^{\ga\gb} \f{\p^2}{\p y^\ga \p \bar y^\gb}
\right )\Big ] C(y,\bar y|\xx,z)=0\,.
\ee

Consider  Eq.~(\ref{dx}) which should reproduce the  rank-two equation
(\ref{unfol2uyR}) at $M=2$. By the substitution
\be
\label{cct}
C(y,\bar y|\xx,z) = \exp ({y_\ga \bar y^\ga})
\,\tilde C(y,\bar y|\xx,z)
\ee
it amounts to
\be
\label{dxt}
\Big [d_\xx -\f{i}{z}d\xx^{\ga\gb}
\f{\p^2}{\p y^\ga \p \bar y^\gb}\Big ] \tilde C(y,\bar y|\xx,z)=0\,.
\ee
Rescaling the variables $y^\ga$ and $\bar y^\dga$
via the substitution
\be
\label{ct}
C(y,\bar y|\xx,z) = z \exp  (y_\ga \bar y^\ga )T(w,
\bar w|\xx,z)\,,
\ee
where the overall factor of $z$ is introduced for the future
convenience and
\be
\label{w}
w^\ga = z^{1/2} y^\ga \q \bar w^\ga =
z^{1/2} \bar y^\ga\,,
\ee
we obtain that $T(w, \bar w|\xx,z)$ satisfies the conformal-invariant rank-two
 unfolded equation
\be
\label{dxt1}
\Big [d_\xx -id\xx^{\ga\gb}
\f{\p^2}{\p w^\ga \p \bar w^\gb}\Big ]  T(w,\bar w|\xx,z)=0\,.
\ee

Substitution of (\ref{ct}) into Eq.~(\ref{dz}) gives
\be
\label{dzt}
\left ( \f{\p}{\p z}
-\half \epsilon^{\ga\gb} \f{\p^2}{\p w^\ga \p\bar w^\gb} \right )
 T(w,\bar w|\xx,z) =0\,.
\ee

Eqs.~(\ref{dxt1}), (\ref{dzt}) are linearized
unfolded equations for 0-forms in $AdS_4$ HS theory in
Poincar{\'e} coordinates. As anticipated, Eq.~(\ref{dxt1}) describes $3d$
conserved currents.  Eq.~(\ref{dzt}) tells us that contractions
$w_\ga \bar w^\ga$ in $T(w,\bar w|\xx,z)$ should carry appropriate powers of $z$.
This conforms to the fact that,
by virtue of the linearized equation (\ref{dxt1}), most of such components
 vanish as a consequence of the conservation equations
for the currents $J$ and $\tilde J$ (\ref{JJ}).
Important exception is provided by $J^{asym}$ (\ref{Jasym}) that describes
the scalar mode of conformal dimension $\Delta=2$
properly accounted by Eq.~(\ref{dzt}) (see also Section \ref{ww}).

\subsection {1-forms}

Using background connection (\ref{w0}) and
 Weyl star product,  we obtain in the sector of HS gauge fields
\be
D_\xx W(y^\pm|\xx,z) =\left (d_\xx +\f{2i}{z}
d\xx^{\ga\gb} y^-_\ga \f{\p}{\p y^{+\gb}}
\right )W(y^\pm|\xx,z)\,,
\ee
\be
D_z W(y^\pm|\xx,z) =\left (d_z -\f{dz}{2z}
\Big (y^+_\ga\f{\p}{\p y^+_\ga}
-y^-_\ga \f{\p}{\p y^-_\ga} \Big )
\right )W(y^\pm|\xx,z)\,.
\ee

Setting
\be
W(y^\pm|\xx,z) = \Omega (v^-, w^+|\xx,z),
\ee
where
\be
\label{v}
v^\pm = z^{-1/2} y^\pm\q w^\pm = z^{1/2} y^\pm\,,
\ee
this gives
\be
D_\xx \Omega(v^-, w^+|\xx,z) =\left (d_\xx +{2i} d\xx^{\ga\gb}
v^-_\ga \f{\p}{\p w^{+\gb}}
\right )\Omega(v^-, w^+|\xx,z)\,,
\ee
\be
D_z \Omega(v^-,w^+|\xx,z) =d_z \Omega(v^-,w^+|\xx,z)\,.
\ee

Now consider Eq.~(\ref{CON12}) starting from  $\xx\,\xx$
sector. Its r.h.s. takes the form
\be
\label{rh}
-\Hh{}^{\ga\gb}\Big ( \eta
\f{\p^2}{\p \overline{w}^{\ga} \p \overline{w}^{\gb}}
{T^{1-i\,i}}(0,\overline{w}\mid \xx,z) + \bar \eta
\f{\p^2}{\p {w}^{\ga} \p {w}^{\gb}}
{T^{i\,1-i}}(w,0\mid \xx,z)\Big )\,,
\ee
where
\be
\Hh^{\ga\gb} = \f{1}{4}
d\xx^{\ga}{}_\gga \wedge d\xx^{\gb \gga}\,.
\ee
Let us stress that explicit dependence on $z$ in Eq.~(\ref{rh})
due to derivatives over $y$ and $\bar y$, and the factors of $z$ in the
frame field (\ref{poframe}) and the definition of
$T$ (\ref{ct}) cancel out.
Using
\be
w_\ga = w^+_\ga + i zv^-_\ga \q \bar w_\ga = i w^+_\ga +z v^-_\ga \,,
\ee
Eq.~(\ref{CON12}) acquires the form
\be
D_\xx \Omega^{jj}_\xx(v^-, w^+|\xx,z) =
\Hh{}^{\ga\gb}
\f{\p^2}{\p {w}^{+\ga} \p {w}^{+\gb}} \big(
{\bar \eta T^{j\,1-j}}(w^+ +iz v^-,0\mid \xx,z)-  \eta
{T^{1-j\,j}}(0,i{w^+} +z v^-\mid \xx,z)\big )\,.
\ee
In the limit $z\to 0$ this gives
\be
\label{dO}
D_\xx \Omega^{ii}_\xx(v^-, w^+|\xx,0) =
\Hh{}^{\ga\gb}
\f{\p^2}{\p {w}^{+\ga} \p {w}^{+\gb}} {\mathcal T}^{ii} (w^+,0 \mid \xx,0)\,,
\ee
where
\be
{\mathcal T}^{jj}(w^+,{w^-} \mid \xx,z) =
{\bar \eta T^{j\,1-j}}(w^+, w^- \mid \xx,z)-
{ \eta T^{1-j\,j}}(-i w^-,{iw^+} \mid \xx,z)\,
\ee
satisfies the rank-two equation (\ref{dxt1})
\be
\label{dxt2}
\Big [d_\xx -id\xx^{\ga\gb}
\f{\p^2}{\p w^{+\ga} \p  w^{-\gb}}\Big ]
 {\mathcal T}^{ii}(w^+, w^-|\xx,z)=0\,.
\ee

Eqs~(\ref{dO}), (\ref{dxt2}) are linearized  unfolded equations
of free $3d$ conformal HS theory that describes
conserved currents ${\mathcal T}^{ii}$ and conformal HS gauge
fields $\Omega_\xx^{ii}$. Being inherited from the nonlinear HS theory in
$AdS_4$, the full boundary theory
should be a  nonlinear  conformal HS gauge theory of
currents ${\mathcal T}^{ii}$ interacting with gauge fields $\Omega^{ii}_\xx$
of Chern-Simons type. We will come back to this issue in Section \ref{n3dconf}.

In  the $z\,\xx$ sector, Eq.~(\ref{CON12}) gives
\bee
\label{dxzO}
&&D_\xx \Omega^{jj}_z(v^-, w^+|\xx,z) + D_z
\Omega^{jj}_\xx (v^-, w^+|\xx,z)=\nn\\
&&\ls \ls =-\f{i}{2}
d\xx{}^{\ga\gb}dz
\f{\p^2}{\p {w}^{+\ga} \p {w}^{+\gb}} \big(
{\bar \eta T^{j\,1-j}}(w^+ +iz v^-,0\mid \xx,z)+
\eta {T^{1-j\,j}}(0,i{w^+} +z v^-\mid \xx,z)\big )\,.
\eee
Eqs.~(\ref{dzt}) and (\ref{dxzO}) determine
$z$--evolution of  $\Omega^{jj}_\xx(v^-, w^+|\xx,z)$ and
${\mathcal T}^{jj}(w^\pm | \xx,z)$. According to
\cite{Akhmedov:1998vf,de Boer:1999xf,Skenderis:2002wp}, supplemented with
nonlinear corrections,
these should be interpreted as renormalization group equations.

\subsection{Weyl, Wick and Fock}
\label{ww}

Let us consider in more detail the relation between the Wick
star-product formalism of Section \ref{ads4cft3} and the
Weyl star-product formalism used in Section \ref{cinf}.

Naively, the boundary limit $z\to 0$ of the map (\ref{Nd}) gives the
identity operator since
\be
\label{z0z}
 f_N(w^\pm|z ) =\exp\left (-2z \epsilon^{\ga\gb}\f{\p^2}{\p w^{-\ga}\p w^{+\gb}}\right )
 f(w^\pm |z)
 \,.
\ee
This is however not  true because of the exponential factors
in (\ref{TN}) and (\ref{ct}) which are singular in the limit $z\to 0$
at $w$ fixed. Moreover, formulas (\ref{TN}) and (\ref{ct}) seemingly
do not match each other because the exponential factors look different
taking into account that
\be
y^-_\ga y^{+\ga} = -\half y_\ga \bar y^\ga\,.
\ee
This is however just the effect of using different star products
in the respective formulas.

Indeed, consider a Weyl star-product element of the form
\be
c(y^\pm) = \exp (-2 y^-_\ga y^{+\ga})\, t(w^\pm)
\ee
which is an analogue of (\ref{ct}). Using (\ref{N}), it easy to see
that
\be
c_N(y^\pm) = \exp (- y^-_\ga y^{+\ga})
\f{1}{\pi^2}\int d^4 u^\pm \exp(-  u^-_\ga u^{+\ga}) t\Big
(\half(w^\pm +z^{1/2} u^\pm)\Big)\,.
\ee
Similarly to (\ref{z0z}),
integration over $u^\pm$ trivializes at $z\to 0$  giving
\be
c_N(y^\pm)\Big |_{z=0} = \exp (- y^-_\ga y^{+\ga}) \, t\Big
(\half w^\pm \Big)\,.
\ee
The exponential factor in this formula just matches
that in (\ref{TN}).

In fact, in the both of star products, the exponentials
\be
\label{fock}
F_N = \exp - y^-_\ga y^{+\ga}\q
F=\exp - 2 y^-_\ga y^{+\ga}
\ee
provide star-product realization of the Fock vacuum that
satisfies
\be
y^-_\ga * F = y^-_\ga \star F_N =0\q
F*y^+_\ga = F_N \star y^+_\ga =0\,.
\ee
Correspondingly, the substitution of the exponential as in
(\ref{TN}) maps Wick star product $\star$ to the operation
$\circ$  that describes action of normal-ordered operators
in the Fock bimodule generated from the Fock vacuum (\ref{fock})
 \be
f(y^-_\gb,y^+_\ga) \circ T(y^\pm) =
f(\f{\p}{\p y^{+\gb}}, y^+_\ga) T(y^\pm)\,,\quad
T(y^\pm)\circ
f(y^-_\gb,y^+_\ga)= f(y^-_\gb,-\f{\p}{\p y^{-\ga}}) T(y^\pm)\,,
\ee
where derivatives $\f{\p}{\p y^{\pm\gb}}$ act  on $T(y^\pm)$.

Note that, in the conformal setup,
 the exponential factor in (\ref{ct})
trivializes for all primary fields
except for $J^{asym}$ (\ref{Jasym})
 since all other primaries depend either only on $y^-$ or only
on $y^+$. In the case of $J^{asym}$, the exponential factor
accounts properly the asymptotic $z$--dependence of $J^{asym}$
in accordance with its conformal dimension $\Delta=2$.

Thus, the Wick  and Weyl star-product descriptions
match at $AdS_4$ infinity.
Nonlocality of the holographic conformal map in the bulk trivializes
at $AdS_4$ infinity, leading to the standard $AdS/CFT$
 prescription where boundary values of the HS gauge fields
in the bulk are identified with sources to conformal operators of
the boundary theory.

\section{Towards nonlinear $3d$ conformal HS theory}
\label{n3dconf}

Although unfolded formulation of the nonlinear $3d$ conformal HS
theory is not yet known,
it can be systematically reconstructed from the $AdS_4$  HS theory
via extension of  analysis of Section \ref{cinf} to the
nonlinear level. Detailed derivation of the nonlinear $3d$ conformal
HS theory will be presented elsewhere. Here we only comment on its general
structure.

Generally, the holographic image of the $AdS_4$ HS theory
should be nonlinear. This is because $AdS_4$  HS gauge connections
contain background and fluctuational parts as two pieces of the
same field. The same should be true in its conformal version.
To be formally consistent,  a system of conformal HS equations
should  be nonlinear with the only exception for the case where r.h.s.
of Eq.~(\ref{dO}) can be zero in all orders. Let us explain this in
some more detail.

To simplify notations, we consider the purely bosonic case with all fields even
in spinor variables $Y^A, Z^A, dZ^A$,
where one can discard the doubling of fields in the full nonlinear system
via truncation of the theory by the projectors (\ref{Pi}). Correspondingly,
in this section we discard the labels
$i,j$ writing $\Omega_{\xx}$ and ${\mathcal T}$ instead
of $\Omega^{ii}_{\xx}$ and ${\mathcal T}^{ii}$.

Let us decompose
\be
\Omega_\xx(v^-, w^+|\xx) = \sum_{n,m=0}^\infty \Omega^{n,m}_\xx(v^-, w^+|\xx)
\q
R_{1\,\xx\xx}:= D_\xx \Omega_\xx = \sum_{n,m=0}^\infty R^{n,m}_{1\,\xx\xx}(v^-, w^+|\xx)
\,,
\ee
\be
A^{n,m}(v^-, w^+|\xx) =
A^{n,m}_{{\ga_1\ldots \ga_n\,\gb_1\ldots \gb_m}}(\xx)
v^{-\ga_1}\ldots v^{-\ga_n} w^{+\gb_1}\ldots w^{+\gb_m}\,.
\ee
Recall that a spin $s$ gauge field is described by
$\Omega^{n,m}_\xx(v^-, w^+|\xx)$ with $n+m=2(s-1)$.
A particularly important role is played
by the fields
\be
\Omega^-_\xx(v^-|\xx) = \sum_{n} \Omega^{n,0}_\xx(v^-, 0|\xx)\,,
\ee
and curvatures
\be
R^+_{\xx\xx}( w^+|\xx) = \sum_{m} R_{1\,\xx\xx}^{0,m}(0, w^+|\xx)\,.
\ee
$\Omega^-_\xx(v^-|\xx)$ and $R^+_{\xx\xx}( w^+|\xx)$ belong to
the subspaces of, respectively, lowest and highest vectors of
the $3d$ conformal subalgebra $sp(4)$ of the $3d$ conformal HS
algebra.

$\Omega^-_\xx(v^-|\xx)$ is the generating function for dynamical
conformal $HS$ gauge fields as can be seen by
virtue of the $\sigma_-$ cohomology analysis with
\be
\sigma_- = e^{\ga\gb}P_{\ga\gb} ={2i} d\xx^{\ga\gb}
v^-_\ga \f{\p}{\p w^{+\gb}}
\,.
\ee
Dynamical fields associated with $H^1(\sigma_-)$ are
rank-$s$ totally symmetric multispinor fields
$\varphi_{\ga_1\ldots \ga_{2s}}(\xx)$  representing
$\Omega^-_\xx(v^-|\xx)$  up to
Lorentz and dilatation  HS gauge shift symmetries
\be
\label{ovar}
\Omega^-_\xx(v^-|\xx) = e^{\ga_1\gb_2}v^{-\ga_3}\ldots v^{-\ga_{2s}}
\varphi_{\ga_1\ldots \ga_{2s}}(\xx)\,.
\ee
 $\varphi_{\ga_1\ldots \ga_{2s}}(\xx)$ provides  $3d$ spinor realization
of traceless conformal HS gauge fields introduced by Fradkin and Tseytlin for
$d=4$ in \cite{conf1}. Having the  gauge transformation law
\be
\delta \varphi_{\ga_1\ldots \ga_{2s}} =  \p_{(\ga_1\ga_2}
\varepsilon_{\ga_3\ldots \ga_{2s})}\,,
\ee
 they are dual to conserved conformal
currents, serving as sources for correlators of currents in the holographic
interpretation.

At the linearized level, $R^+_{\xx\xx}(w^+|\xx)$
is the part of the linearized conformal HS curvature that
contains nontrivial gauge invariant combinations of
derivatives of the dynamical fields $\varphi_{\ga_1\ldots \ga_{2s}}$.
Namely, the $\sigma_-$ cohomology analysis shows that the conditions
\be
\label{R0}
R_{1\,\xx\xx}^{n,m}=0 \q n>0
\ee
give algebraic constraints which
express  fields $\Omega^{n,m}$ via order--$m$ derivatives  of
the dynamical fields, imposing no restrictions
on the latter. Although most of components of
$R^+_{\xx\xx}(w^+|\xx)$ vanish by virtue of Bianchi identities applied
to (\ref{R0}), some may remain nonzero. These are just those
parametrized by the 0-forms ${\mathcal T}({w^+,0} \mid \xx)$ on the
r.h.s. of the unfolded equations (\ref{dO}) (discarding indices $i$). In fact,
${\mathcal T}({w^+} ,0\mid \xx)$ just represents $H^2(\sigma_-)$.

Important property of the unfolded system (\ref{dO}), (\ref{dxt2})
with currents ${\mathcal T}({w^+,w^-} \mid \xx)$ treated as independent $3d$ fields
is that it is off-shell. This means that the system (\ref{dO}), (\ref{dxt2}) expresses
up to gauge transformations all
fields $\Omega_\xx(v^-, w^+|\xx)$ and ${\mathcal T}({w^+,w^-} \mid \xx)$ via derivatives
of $\varphi_{\ga_1\ldots \ga_{2s}}$  imposing no restrictions
on the latter. In particular, this means that Eqs.~(\ref{dxt2})
are consequences of Eqs.~(\ref{dO}) supplemented with constraints which
express ${\mathcal T}({w^+,w^-} \mid \xx)$ via derivatives
of ${\mathcal T}({w^+,0} \mid \xx)$.
In this setup, the current conservation equation
\be
\label{cc}
\f{\p}{\p \xx^{\ga\gb}}\f{\p^2}{\p w^+_\ga \p w^+_\gb} {\mathcal T}(w^+,0|\xx) =0
\ee
follows from the expression for ${\mathcal T}(w^+,0|\xx)$ in terms of derivatives
of $\varphi_{\ga_1\ldots \ga_{2s}}$ by virtue of Eq.~(\ref{dxt2}). Other way around,
given conserved current ${\mathcal T}({w^+,0} \mid \xx)$, Eq.~(\ref{dxt2}) determines
dynamical fields $\varphi_{\ga_1\ldots \ga_{2s}}$ in terms of
${\mathcal T}({w^+,0} \mid \xx)$ up to a gauge transformation.

That the $3d$ system (\ref{dO}), (\ref{dxt2}) is off-shell means that
unrestricted  $\varphi_{\ga_1\ldots \ga_{2s}}$
can be interpreted as arbitrary boundary values of the bulk HS gauge fields.
It should be stressed that, being off-shell in $d=3$, the system
(\ref{dO}), (\ref{dxt2}) becomes on-shell in a larger space like $AdS_4$ or
 a  space with additional spinor coordinates
of  Section \ref{ferext}. As discussed in Section \ref{Duality},
this is crucial for holographic interpretation of the theory.

It should be stressed that unfolded dynamics properly accounts asymptotic
behavior of relativistic fields in $AdS$. We have seen this already for
scalar field in $AdS_4$ where unfolded equations reproduces
two asymptotic behaviors with $\Delta=1$ and $\Delta=2$. Analysis of this
section extends this observation to any spin. Indeed, {}Eq.~(\ref{DNT}) shows
that conserved currents have canonical conformal dimension $s+1$ and hence
asymptotic behavior $z^{s+1}$. {}From Eq.~(\ref{dO}) it follows that $\Omega(0,w^+)$
has asymptotic behavior $z^{s-1}$, which is in agreement with (\ref{add}),
taking into account  the fact (\ref{1f}) that the total number of spinor indices
carried by a spin-$s$ connection is $2(s-1)$.  From Eq.~(\ref{v})
it follows that $\Omega(v^-,0)$ has asymptotic behavior $z^{1-s}$ which is again
in agreement with (\ref{add}). Since the background frame field in (\ref{ovar})
contains the factor of $z^{-1}$,
$\varphi_{\ga_1\ldots\ga_{2s}}$
has asymptotic behavior $z^{2-s}$ which is just the correct behavior of the
boundary source.

Setting to zero ${\mathcal T}({w^+} ,0\mid \xx)$  imposes differential equations on the
dynamical fields $\varphi_{\ga_1\ldots \ga_{2s}}$. Since Eq.~(\ref{R0}) at
${\mathcal T}({w^+} ,0\mid \xx)=0$ is the linearized flatness condition, in this
case the HS gauge fields become pure gauge. For nonzero
${\mathcal T}({w^+} ,0\mid \xx)$, conformal HS gauge fields are nontrivial.
To see whether or not the theory can remain free beyond the linearized
approximation one has to check whether or not the condition
$R_{1\,\xx\xx}(v^-,w^+\mid \xx)=0$ is consistent with the
full nonlinear HS equations in $AdS_4$. As discussed in more detail in
Section \ref{bound}, this indeed turns out to be possible for two particular truncations
of HS models: one for $A$--model and another  for $B$-model. Correspondingly,
the related truncated HS theories turn out to be holographically dual to free
boundary bosonic  and fermionic  theories in agreement
with Klebanov-Polyakov \cite{Klebanov:2002ja} and Sezgin-Sundell
\cite{Sezgin:2003pt} conjectures. This conclusion is also in agreement with
Maldacena-Zhiboedov theorem \cite{Maldacena:2011jn} because in these  cases
$3d$ conformal HS gauge fields decouple from the $3d$ currents. However, beyond
these two  cases, the boundary dual of  $AdS_4$ HS theory is
nonlinear.

One reason why the boundary theory should be nonlinear is that the
conformal HS curvatures inherited from $AdS_4$ HS theory
\be
\label{rcon}
R_{\xx\xx}(v^-,w^+\mid \xx) = d_\xx \Omega_\xx(v^-,w^+\mid \xx) +
\Omega_\xx(v^-,w^+\mid \xx)\star \Omega_\xx(v^-,w^+\mid \xx)\,
\ee
are nonlinear. Here it is crucial that the rescalings (\ref{w})
and (\ref{v}) have opposite scalings in the radial coordinate $z$
so that $v^-$ and $w^+$ obey $z$--independent commutation
relations
\be
[v^-_\ga\,,w^{+\gb}]_\star =\delta_\ga^\gb\,.
\ee

Analogously, l.h.s. of Eq.~(\ref{dxt2}) deforms
to the covariant derivative in the
 twisted adjoint representation of the non-Abelian $3d$ conformal HS algebra.
Since the substitution (\ref{TN}) for the current maps
Wick star product to  Fock product,
nonlinear extension of Eq.~(\ref{dxt2})
starts from the twisted adjoint covariant derivative
\be
\label{twj}
\tilde D T(w^\pm|x)= dT(w^\pm|x) +\Omega
(\f{\p}{\p w^{+\gb}}, w^+_\ga) T(w^\pm|x) -
T(w^\pm|x)  \Omega (-i\f{\p}{\p w^{-\ga}},
-i w^-|x)\,,
\ee
where we used (\ref{pm}) and that
$\Omega(-v^-,- w^+)=\Omega(v^-, w^+)$ for bosons.

It should be stressed that  nonlinear terms in Eqs.~(\ref{rcon}) and (\ref{twj}) are
$z$--independent, hence fully reproducing the non-Abelian structure of $3d$ conformal
HS theory in the $z\to 0$ limit. The nonlinear deformation (\ref{twj}) of the
twisted adjoint covariant derivative implies a nonlinear deformation of the
rank-two unfolded  equation (\ref{dxt2}) which, in turn, implies
a nonlinear deformation of the conventional current conservation condition,
hence not respecting conditions of  Maldacena-Zhiboedov theorem \cite{Maldacena:2011jn}.

The nonlinear deformation due to non-Abelian HS algebra is
just a first step requiring further ${\mathcal T}$-dependent nonlinear deformation of
Eqs.~(\ref{dO}) and (\ref{dxt2}). Similarly to \cite{Ann} one can
search these corrections perturbatively
in powers of ${\mathcal T}$.  However,  straightforward analysis is
not simple. It seems more promising to try to guess
a closed form of yet unknown nonlinear $3d$ HS conformal
unfolded system as it was guessed for the $AdS_4$ HS system  in
\cite{con,more}. We plan to consider this problem elsewhere.

It may look a bit peculiar that the variables $v^+$ and $v^-$
appear asymmetrically in Eqs.~(\ref{dO}), (\ref{dxt2}). This happens because of choosing
a particular $y^+ y^-$ Wick ordering in the star product
(\ref{Wick}) or, alternatively, a particular form of connection (\ref{w0}). Choosing the $y^- y^+$
ordering exchanges the roles of $y^+$ and $y^-$. Full nonlinear conformal HS theory is
expected to describe both of these sectors on equal footing.

Another comment is that in the realization of conformal HS theory
described so far the $3d$ conformal currents appear as independent
dynamical objects whose properties are determined by the equations of the
$3d$ conformal HS theory itself.
This model does not capture  $3d$ conformal
scalar and spinor fields $\Phi^i(w^+|x)$ from which the currents
${\mathcal T}(w^\pm|x)$ can be built. We believe that this ingredient can
also be incorporated into  $3d$ conformal HS theory.
Indeed, as shown in \cite{ShV},  $3d$ conformal
scalar and spinor can be described as fields $|\Phi^i(w^+|x)\rangle $
valued in the Fock module of the $3d$ conformal HS algebra.
In these terms,
free field equations for $3d$ conformal fields have the form
\be
(d +\Omega_0 (v^-, w^+|\xx))\star \Phi^i(w^+|\xx)\star F =0\,.
\ee
An interesting problem for the future is to find a nonlinear
$3d$ conformal HS theory for the full
system of fields $\Omega$, ${\mathcal T}$ and $\Phi$, which relates
${\mathcal T}$ to proper bilinear combinations of $\Phi$.
Solution of this problem should clarify explicit relation of our construction
to (generalized) boundary $\sigma$-model constructions of
\cite{Klebanov:2002ja,Sezgin:2003pt}.

\section{Boundary conditions, reductions and $AdS$ doubling }
\label{bound}
Standard $AdS/CFT$ correspondence assumes  certain boundary conditions
at infinity. In terms of $AdS_4$ HS Weyl forms they relate
${ T^{j\,1-j}}(w^+, 0 \mid \xx,0)$ and ${ T^{1-j\,j}}(0,{iw^+} \mid \xx,0)$.
Let
\be
\label{a}
{\mathcal A}^{jj}(w^+,{w^-} \mid \xx ) =
{ T^{j\,1-j}}(w^+, w^- \mid \xx,0)-
{T^{1-j\,j}}(-i w^-,{iw^+} \mid \xx,0)\,,
\ee
\be
\label{b}
{\mathcal B}^{jj}(w^+,{w^-} \mid \xx ) =
{ T^{j\,1-j}}(w^+, w^- \mid \xx,0)+
{ T^{1-j\,j}}(-i w^-,{iw^+} \mid \xx,0)\,.
\ee

Conditions
\be
\label{dir}
{\mathcal A}^{jj}(w^+,{w^-} \mid \xx ) =0
\ee
 and
\be
\label{neum}
{\mathcal B}^{jj}(w^+,{w^-} \mid \xx ) =0
\ee
will be called \Di\, and \Ne, respectively.
We observe that that r.h.s. of Eq.~(\ref{dO}) is zero for the \Di\, boundary
conditions in the $A$-model
($\eta=1$) and for the \Ne\, conditions in the $B$-model
($\eta=i$). On the other hand, r.h.s. of Eq.~(\ref{dxzO}), that determines
$z$-evolution of the HS connection, is zero for the \Ne\,
conditions in the $A$-model and for the \Di\, conditions in the
$B$-model. This suggests that the latter boundary conditions correspond to
IR fixed points of the model.

${\mathcal A}^{jj}$ and ${\mathcal B}^{jj}$ describe independent combinations
of $T^{j,1-j}$, \ie each of ${\mathcal A}$ or ${\mathcal B}$
conditions leaves some components of $T^{j,1-j}$ nonzero. For any other choice of
relative coefficients on the r.h.s.s of Eqs.~(\ref{a}),
(\ref{b}), the corresponding conditions would be too strong, implying
$T^{j,1-j} =0$.

The bosonic model where all fermions are zero decomposes into two
independent systems projected out by the projectors $\Pi_\pm$ (\ref{Pi}). The currents,
that contribute to the $\Pi_+$ model, are
\be
{\mathcal A}_{+}(w^+,{w^-} \mid \xx ) =\sum_{j=0,1}
\Big ( { T^{j\,1-j}}(w^+, w^- \mid \xx,0)-
{T^{1-j\,j}}(-i w^-,{iw^+} \mid \xx,0)\Big )\,,
\ee
\be
{\mathcal B}_{+}(w^+,{w^-} \mid \xx ) =\sum_{j=0,1}\Big (
{ T^{j\,1-j}}(w^+, w^- \mid \xx,0)+
{ T^{1-j\,j}}(-i w^-,{iw^+} \mid \xx,0)\Big )\,.
\ee
In this case,  \Di\, condition implies that the current $J^{asym}(X)$
(\ref{Jasym}) is zero while $J^{sym}(0\mid  X)$ remains free.
\Ne\, condition implies that $J^{sym}(0\mid  X)=0$ and $J^{asym}(X)$
remains free. Hence \Di\, and \Ne\, boundary conditions just distinguish
between two scalar currents that have different
conformal dimensions, namely, $\Delta(J^{sym}) =1$ and $\Delta(J^{asym})=2$.

In the  $\Pi_-$ model, ${\mathcal A}$ and  ${\mathcal B}$
conditions have opposite effect. Namely,  \Di\, condition implies that the current
$J^{sym}(X)$ is zero and $J^{asym}(0\mid  X)$ remains free while
\Ne\, condition implies that $J^{asym}(0\mid  X)=0$ and $J^{sym}(X)$
remains free. This conclusion is in agreement with supersymmetry
of the model in presence of fermions: for any type of boundary conditions,
the supersymmetric model will contain both types of scalars, one in the $\Pi_+$
sector and another in the $\Pi_-$ sector.
  \Di\, and \Ne\, boundary conditions  extend two different
 types of boundary conditions for scalar currents to currents of all spins.

If the r.h.s. of Eq.~(\ref{dO}) is zero, it becomes the flatness
condition for boundary HS gauge fields. In the gauge where nonzero
boundary HS gauge fields belong to the conformal algebra $sp(4)$,
the resulting theory  describes free unfolded equations on boundary currents in some
conformally flat background.
In agreement with Klebanov-Polyakov \cite{Klebanov:2002ja} and
Sezgin-Sundell \cite{Sezgin:2003pt} conjectures these two particular cases correspond
 to the free boundary models of conformal scalar or
spinor in $\Pi_+ A$ and $\Pi_- B$  or $\Pi_+ B$ and $\Pi_- A$-models, respectively.

In fact, the free boundary theories are dual to  truncations of the full
nonlinear HS theories in $AdS_4$ by the parity automorphism ${ P}$ that exchanges
dotted and undotted spinors. Indeed, as observed in
\cite{Sezgin:2003pt}, the nonlinear HS equations (\ref{dW})-(\ref{SS}) are $P$ invariant
provided that ${ P}(B)=B$ in the $A$-model or ${ P}(B)=-B$ in the $B$-model.

There is however an interesting and important subtlety in this consideration.
Indeed, so defined $P$  describes the reflection
$z\to-z$ of the coordinate $z$ as introduced in (\ref{xz}). Hence, to apply
 ${ P}$,  one  Poincar{\'e} chart of $AdS_4$  has to be supplemented with another
one with negative $z$ to allow
\be
\label{pz}
P(z) = - z\,.
\ee

In fact, in our construction it is important that $AdS_4$ is doubled to contain
two  Poincar{\'e} charts related by $P$.
Although, geometrically, $P$ leaves  $AdS_4$ invariant, the effect of this doubling
is nontrivial because extension of solutions from one chart to another is not
necessarily $P$-invariant. For example, this is the case in non $P$-invariant
HS theories with $\eta^2\neq\pm 1$. On the other hand,
no boundary conditions at $z=0$ should be imposed to define the action
as  integral over the doubled $AdS$ space-time. In this
setup holographic duality  relates a bulk theory in the doubled
$AdS$ space to the ``boundary theory" where all possible types of boundary fields
$\phi_{bound}(\xx)$ contribute. In the unfolded dynamics approach,
values of  $\phi_{bound}(\xx)$ at  $z=0$ reconstruct all fields in the
(doubled) bulk and hence values of the respective action functionals $S(\phi_{bound})$.
Note that in this respect the situation with the surface $z=0$ in the
doubled bulk space is analogous to
that with a regular $3d$ surface $\Sigma$ inside bulk as discussed in Section
\ref{ads4cft3}.

We believe that the doubled bulk $AdS/CFT$ setup, which follows naturally from unfolded
dynamics, has  general  applicability for HS theories and beyond.
The important issue of anomalies also fits naturally this
 problem setting as we discuss briefly in Section \ref{actions}.

In terms of elementary oscillators of the $AdS_4$ HS theory, $P$  acts as follows
\be
\label{py}
P(y_\ga)= \bar y_\ga\q P(\bar y_\ga ) = -y_\ga
\ee
which is equivalent to
\be
P(y^\pm_\ga ) =\pm i y^\pm_\ga\,.
\ee
Being defined in such a way that it maps $y^+$ and $y^-$ to themselves, $P$
is not involutive. Namely,
$
P^2 = F,
$
$ F^2=Id,
$
where $F$ is the boson-fermion automorphism that changes a sign of fermions. Although,
naively, this property obstructs consistent $P$-reduction of the HS theory
in presence of fermions, this is not the case. Remarkably, this is just what
doctor ordered to cure
additional factors of $i$ that appear due to the $z\to -z$ reflection of
the factor of $z^{1/2}$ in the rescaling (\ref{w}), so that (\ref{py}) is replaced by
\be
P(w^\pm_\ga ) =\pm  w^\pm_\ga\,.
\ee

So defined $P$  admits two  extensions $P_\pm$  to the full
nonlinear HS system
\be
P_\pm(z_\ga)=- \bar z_\ga\q P_\pm(\bar z_\ga ) = z_\ga\q
P_\pm(dz^\ga)=- \bar d\bar z^\ga\q P_\pm(\bar dz^\ga ) = dz^\ga\,,
\ee
\be
 P_\pm(k)=\pm \bar k
\q P_\pm(\bar k) = \pm k\,.
\ee
(Spinor coordinates $z_\ga$ and $\bar z_\ga$ should not be confused with
the radial coordinate $z$ of $AdS_4$.)  $P_\pm$
leave invariant all nonlinear equations except for Eq.~(\ref{SS}) which is not invariant
for general $\eta$. However, Eq.~(\ref{SS}) is invariant under $P_+$ and $P_-$,
in the cases of $A$ and $B$ models, respectively. This allows us to truncate
nonlinear $A$ and $B$-models  in $AdS_4$ by  the conditions
\be
\label{Pc}
{ P_\pm} W= W\q { P_\pm } S= S\q { P_\pm } B= B\,.
\ee
Associated boundary theories are $A$ and $B$--models with, respectively, ${\mathcal A}$ and
${\mathcal B}$ boundary conditions at the linearized level. Since in these cases
boundary currents decouple from $3d$ conformal HS gauge fields
at the linearized level and since the conditions (\ref{Pc}) are consistent in all
orders, the corresponding truncations  of the bulk HS theory should correspond to the
free boundary theories in all orders.

Other models and/or boundary conditions do
not correspond to any consistent truncation of the full bulk theory.
Even if the boundary conditions  were imposed in such a way
that the r.h.s. of  Eq.~(\ref{dO}) be zero in the lowest order, it will
acquire non-zero higher-order corrections from the full nonlinear system.
From the perspective of \cite{Maldacena:2012sf}, these cases correspond to broken
HS symmetry because the   current conservation equations are
deformed by nonlinear corrections, \ie the currents are not conserved in the conventional
sense. {}From the  bulk HS theory perspective this is the effect of a nonlinear
deformation of the HS gauge transformation law rather than breaking of HS symmetry.

To summarize,
except for the particular cases of \Di\, boundary condition in the $A$-model
 and \Ne\, boundary condition in the $B$-model,
all other possibilities correspond to  nonlinear boundary
conformal HS theories where boundary conformal HS gauge fields are sourced by
boundary currents. This leads to  fully nonlinear boundary theories
where currents interact via Chern-Simons type boundary conformal HS gauge fields.
In particular, this happens for all HS theories with $\eta^2\neq \pm 1$.

The holographic duality described in this paper works for any
coupling constant in the HS theory, hence not referring to the
 $N\to\infty$ limit. In this respect it extends the Klebanov-Polyakov-Sezgin-Sundell
conjecture on the critical $O(N)$ and Gross-Neveu models to
finite $N$. Beyond the $N\to\infty$ limit, $AdS_4$ HS theory is shown
to be dual to a nonlinear theory that describes HS interactions of
boundary currents via $3d$ conformal HS gauge fields. It remains to be
seen what is the relation of this boundary HS theory to critical
$O(N)$ and Gross-Neveu models as well as to the models with arbitrary $\eta$
discussed in \cite{Giombi:2011kc,Aharony:2011jz}.

\section{$AdS_3/CFT_2$}
\label{32}

\subsection{$AdS_3$ description}
$AdS_3/CFT_2$ correspondence in HS theories has been extensively
studied in \cite{Henneaux:2010xg,Campoleoni:2010zq,Gaberdiel:2010pz,
 Gaberdiel:2011wb,Ahn:2011pv,Gaberdiel:2011zw,Chang:2011mz,
 Gaberdiel:2011nt,Campoleoni:2011hg,Kraus:2011ds,Ammon:2011ua,Gaberdiel:2012yb,Candu:2012jq}.
 It is interesting to reconsider this problem  along the lines of
 the $AdS_4/CFT_3$ analysis of previous sections. For details of the
 nonlinear $AdS_3$  HS theory we refer the reader to \cite{prok} (see also
\cite{Vasiliev:1999ba}).  Below we only need the linearized construction.

The $AdS_3$ algebra is semisimple: $o(2,2)\sim sp(2;R)\oplus sp(2;R)$
with the diagonal subalgebra $sp(2;R)\sim o(2,1)$
as Lorentz algebra. A particularly
useful realization of  $AdS_3$ generators is
\be
\label{al}
   L_{\ga\gb}=\frac1{4i}\{\hat{y}_\ga,
    \hat{y}_\gb\}\,,\qquad \,
   P_{\ga\gb}=\frac1{4i}\{\hat{y}_\ga,
    \hat{y}_\gb\}\psi  \,
\ee
with the generating elements $\hat{y}_\ga$ and $\psi$ obeying the relations
$[\hat{y}_\alpha , \hat{y}_\beta ] = 2i \epsilon_{\alpha\beta}$,
$\epsilon_{\alpha\beta}=-\epsilon_{\beta\alpha}$, $\epsilon_{12}=1$
and
\be
\psi^2=1\,,\qquad [\psi ,\hat{y}_\alpha ]=0\,.
\ee
$\Pi_\pm = \frac{1}{2} (1\pm \psi )$ are
projectors to the simple components of $o(2,2)$.

In \cite{BF,BBS,H,Aq,FLU} it was shown that there exists a
one-parametric class of infinite-dimensional algebras
 $ hs(2;\nu )$ ($\nu$ is an arbitrary real parameter),
all containing $sp(2)$ as a subalgebra. This allows one to
define a class of HS algebras
$
g=hs(2;\nu )\oplus hs(2;\nu )
$
which admit a useful realization in terms of deformed oscillators.

Consider associative algebra
$Aq(2;\nu )$ \cite{Aq}
of elements of the form
\be
\label{sel}
f(\ty,k )=
\sum^\infty_{n=0}\sum_{A=0,1}
\!
\frac{1}{n!}
 f^{A\,\alpha_1\ldots\alpha_n}(k)^A
 \ty_{\alpha_1}\ldots \ty_{\alpha_n}\,
\ee
under condition that the coefficients
$ f^{A\,\alpha_1\ldots\alpha_n}$ are symmetric with respect to
 the indices
$\ga_j $ and that the generating elements $\ty_\ga$ obey
\bee
\label{modosc}
[\ty_\ga ,\ty_\gb ]=2i
\epsilon_{\ga\gb} (1+\nu k)\,,\quad k\ty_\ga
=-\ty_\ga k\,,
\quad k^2 =1\,,
\eee where $\nu $ is an arbitrary constant (central element).
In other words,
$Aq(2;\nu )$ is the enveloping algebra of the relations
(\ref{modosc}) often called deformed oscillator algebra.

Its important property  is that, for all $\nu$,
\be
\label{lorq}
T_{\ga\gb}=\frac{1}{4i}\{\ty_\ga \,,\ty_\gb \}
\ee
obey $sp(2)$ commutation relations,  rotating $\ty_\ga$ as a $sp(2)$ vector
\bee
\label{q1com}
[T_{\ga \gb},
T_{\gga \eta}]\!=\! \epsilon_{\ga\gga} T_{\gb\eta}\!+\!
\epsilon_{\gb\gga} T_{\ga\eta}\!+\!
\epsilon_{\ga\eta} T_{\gb\gga}\!+\! \epsilon_{\gb\eta} T_{\ga\gga}\,,
\eee
\bee
\label{q2com}
[T_{\ga\gb} , \ty_{\gga }]
\!=\! \epsilon_{\ga\gga}\ty_{\gb} \!+\! \epsilon_{\gb\gga}\ty_{\ga}.
\eee

Deformed oscillators
were originally discovered by Wigner \cite{wig}
who addressed the question whether it is possible to modify
the commutation relations for usual oscillators
$a^\pm$ in such a way
that the basic commutation relations $[H, a^\pm] =\pm a^\pm $,
$H=\frac{1}{2} \{a^+ ,a^- \}$ remain
intact. By analyzing this problem in the Fock-type space
Wigner found a one-parametric deformation of the standard commutation
relations which gives a particular representation of the
commutation relations (\ref{modosc})  with the identification
$a^+ =\ty_1$,
$a^- =\frac{1}{2i} \ty_2$, $H=T_{12}$ and $k=(-1)^N$ where $N$ is the
particle number operator. These commutation relations were
discussed later on by many authors (see, e.g.,~\cite{yang,deser,mukunda}).

According to (\ref{lorq}) and (\ref{q2com}),
the $sp(2)$  generated by $T_{\ga\gb}$
extends to $osp(1,2)$ via identification of  supergenerators with
$\ty_\ga$. The quadratic Casimir operator of
$osp(1,2)$
\be
C_2 = -\frac{1}{2}T_{\ga\gb}T^{\ga\gb} -\frac{i}{4}\ty_\ga \ty^\ga\,,
\ee
is
\be
C_2 =-
\frac{1}{4} (1- \nu^2)\,.
\ee

Thus, $Aq(2,\nu )$  is isomorphic to
$U(osp(1,2))/I_{C_2 + \frac{1}{4} (1- \nu^2)}$, where
the ideal $I_{C_2 + \mu}$ consists of elements proportional
to $C_2 + \mu$, $\mu\in \mathbb{C}$. This  has  a number of
 important consequences. For example, any module of
$osp(1,2)$ with $C_2 =  -\frac{1}{4} (1- \nu^2)$ forms a module of
$Aq(2,\nu )$ ($\nu \neq 0$) and vice versa.
In particular this is the case for finite-dimensional
modules corresponding to  $\nu =2l+1$, $l\in
{\bf Z}$ with $C_2 = l(l+1)$.

The even subalgebra of $Aq(2;\nu )$
spanned by $f(\ty,k)$  (\ref{sel}) obeying $f(\ty,k)=f(-\ty,k)$
decomposes into direct sum of two subalgebras
$Aq^E_\pm (2;\nu )$ spanned by the elements $\Pi_\pm f(\ty,k)$ with
$f(-\ty,k)=f(\ty,k)$, $\Pi_\pm = \frac{1}{2} (1\pm k)$. These algebras
are  isomorphic to
$U(sp(2))/I_{C_2 + \frac{3\pm 2\nu -\nu^2}{4}} $,
where
$C_2 = -\frac{1}{2}T_{\ga\gb}T^{\ga\gb}$ is the quadratic
Casimir operator of $sp(2)$, and can be interpreted
as (infinite-dimensional) algebras interpolating between  ordinary
finite-dimensional matrix algebras as discussed in \cite{BF,FLU}.

Algebra $o(2,2)\sim sp(2)\oplus sp(2)$ can be
spanned by $\psi$--dependent bilinears of the oscillators $\hat{y}$.
Its HS extension results from allowing
all powers of $\hat{y}$. HS gauge fields
are
\be
\label{ghs}
w(\ty,\psi,k|x)
  \!\! =  \!\!\!\! \sum^\infty_
    {A,B=0,1;\;n=0 }
    \frac{1}{n!}  w^{A\,B\ga_1\ldots\ga_n}(x) k^A \psi^B
    \ty_{\ga_1}\ldots \ty_{\ga_n}\,.
\ee
 $w^{A\,B\ga_1\ldots\ga_n}(x)$ describe
$3d$ HS gauge fields of spin $\half n$. HS curvatures have standard form
\be
\label{cur}
   R(\ty,\psi,k|x) = dw(\ty,\psi,k|x) +
   w(\ty,\psi,k|x) w(\ty,\psi,k|x) \,.
\ee
(This construction for ordinary (\ie $\nu$=0) oscillators was
suggested in \cite{bl}.) The labels $A=0,1$ and $B=0,1$ play
different roles. $A$ describes the doubling of all fields
as a consequence of $N=2$ supersymmetry in the theory.
This doubling can  be avoided in an appropriately
truncated theory \cite{prok}. $B$ distinguishes
between the Lorentz--like $(B=0)$ and frame--like
 $(B=1)$ fields.

The $3d$ linearized system is simpler than the $4d$
one because, analogously to  $3d$ gravity \cite{d3gr,d3grw},
$3d$ HS fields do not propagate being of Chern-Simons type.
Equivalent statement is that $3d$ HS fields admit no HS Weyl
tensors. Consequently, the $3d$ Central On-Mass-Shell
Theorem has the form
\be
 R_1(\hat{y} ,\psi ,k|x):= dw(\ty,\psi,k|x) +
   w_0 (\ty,\psi,k|x)  w(\ty,\psi,k|x) +
    w(\ty,\psi,k|x)  w_0(\ty,\psi,k|x)
 =0 \,,
\ee
\be
\label{d3l}
 {\cal D}_0 C(\hat{y}, \psi ,k |x) :=d C(\hat{y}, \psi ,k |x)
+ w_0(\hat{y}, \psi ,k |x)C(\hat{y}, \psi ,k |x)
-C(\hat{y}, \psi ,k |x)w_0(\hat{y},- \psi ,k |x)=0
\,,
\ee
where $w_0(\hat{y} ,\psi ,k|x)$ is some $AdS_3$ flat connection.

As shown in  \cite{BPV}, in the sector of 0-forms,
(\ref{d3l}) describes four massive scalars,
$C(\hat{y},\psi ,k|x)=C(-\hat{y},\psi ,k |x)$,
and four massive spinors,
$C(\hat{y},\psi ,k|x)=-C(-\hat{y},\psi ,k |x)$,
arranged into $N=2$ $3d$ hypermultiplets.  Masses $M$ of matter fields
are expressed in terms of $\lambda$ and $\nu$ as follows \cite{BPV}
\be
\label{M}
  M^2_\pm =\gl^2\frac{\nu(\nu\mp 2)}2 \,
\ee
for bosons, and
\be
\label{M f}
  M^2_\pm =\gl^2\frac{\nu^2}2 \,
\ee
for fermions. Here $-\lambda^2$ is the cosmological constant of $AdS_3$.
The signs ``$\pm$'' refer to the projections
$
   C^\pm=\Pi_\pm C$,  $\Pi_\pm=\frac{1\pm k}2
$.
Doubling  of fields of the same mass is due to $\psi$ ($\psi^2 =1$) while
that with mass splitting in the bosonic sector, is due to $k$.
Component form of the covariant constancy conditions
(\ref{d3l}) was originally found in \cite{BPV}
(see also \cite{Vasiliev:1999ba}).

\subsection{$CFT_2$ description}

Analysis of conformal version of
$AdS_3$ HS theory is to some extent parallel to the $AdS_4$
case. Radial coordinate $z$ is identified with
$
z=x^{12}
$
while the boundary coordinates are $\xx=x^{11}$ and
$\tilde\xx=x^{22}$
\be
x^{\ga\gb }=(\xx^{\ga\gb}, \sigma_1^{\ga\gb} z)\q
\sigma_{1\,{\ga\gb}}\xx^{\ga\gb}=0\q
\sigma_1^{12}=\sigma_1^{\,21}=1\,.
\ee
The $3d$ conformal algebra $o(2,2)$ as well as its HS
extension decomposes into direct sum of two subalgebras
$
o(2,2)=sp(2)\oplus \widetilde{sp}(2)
$
\be
sp(2): T_{\ga\gb}=\f{1}{8i}(1+\psi)(\hat y_\ga \hat y_\gb+
\hat y_\gb \hat y_\ga)\q
\widetilde{sp}(2):
\tilde T_{\ga\gb}=\f{1}{8i} (1-\psi)(\hat y_\ga \hat y_\gb+ \hat
y_\gb \hat y_\ga)\,.
\ee
Lorentz and dilatation generators are defined by the relations
\be
L-D = \half \sigma_1^{\ga\gb} T_{\ga\gb}\q
L+D = \half \sigma_1^{\ga\gb} \tilde T_{\ga\gb}\,.
\ee
{}From here it follows that
\be
[D\,, T_{22}]= - T_{22}\q [D\,, T_{11}]=  T_{11}\q
[D\,, T_{12}]= 0\,,
\ee
\be
[D\,, \tilde T_{22}]=  \tilde T_{22}\q [D\,, \tilde T_{11}]= -
\tilde T_{11}\q [D\,, \tilde T_{12}]= 0\,.
\ee

In accordance with (\ref{D}) we  set
\be
P=T_{22}\q \tilde P = \tilde T_{11}\,.
\ee
Poincar{\'e} foliated flat connection (\ref{fol}) is
\be
W_0 = z^{-1} \big ( d\xx  P+d\tilde \xx \tilde P - dz D \big ) \,.
\ee

Manifest conformal invariance is achieved via
transition from the Weyl star product in $AdS_3$ setup to
the Fock bimodule realization with the Fock vacua $\F_\pm$
that satisfy
\be
y_1 \circ \F_\pm = \F_\pm \circ y_2 =0\q k\F_\pm = \F_\pm k =
\pm \F_\pm\,.
\ee
An element $F_\pm(y)$ of the Fock bimodule results from the
vacuum $\F_\pm$ via action of functions of $y_2$ from the
left and functions of $y_1$ from the right. This gives
\be
y_1\circ F_\pm (y)= 2i \D_1 F_\pm (y)\q y_2\circ
F_\pm(y) = y_2 F_\pm(y)\,,
\ee
\be
F_\pm(y) \circ  y_1= F_\pm(y)   y_1\q F_\pm(y) \circ  y_2=2i
 F_\pm (y)\overleftarrow{\D_2}\,,
\ee
where
\be
\D_1 F_\pm (y_1,y_2) = \f{\p}{\p y^1}F_\pm (y_1,y_2)
\pm \f{\nu}{2y_2}(F_\pm (y_1,y_2)
-F_\pm (y_1, -y_2))\,,
\ee
\be
 F_\pm (y_1,y_2)\overleftarrow{\D_2} =
 \f{\p}{\p y^2}F_\pm (y_1,y_2)
\pm \f{\nu}{2y_1}(F_\pm (y_1,y_2)
-F_\pm (- y_1,  y_2))\,.
\ee
Note that $\D_1$ and $\overleftarrow{\D_2}$ are so-called Dunkl
derivatives \cite{Dunkl} of the two-body Calogero model.

In this setup, the system becomes manifestly conformal with
homogeneous polynomials of $y_{\ga}$ carrying definite
conformal dimensions in the adjoint
\be
[D\,,A_\pm(y)] =\half \left (y^2 \f{\p}{\p y^2}-
y^1 \f{\p}{\p y^1}\right ) A_\pm(y)\q
[D\,,\tilde A_\pm(y)] =
\half \left (y^1 \f{\p}{\p y^1}-
y^2 \f{\p}{\p y^2}\right ) \tilde A_\pm(y)
\ee
and twisted adjoint representation
\be
D( F_\pm(y))  =\half \left (\Big (y^\ga \f{\p}{\p y^\ga}
 +2(1 \pm \nu )  \Big ) F_\pm(y) \mp\nu  ( F_\pm(-y_1,y_2)+
F_\pm(y_1,-y_2)) \right )\,,
\ee
\be
D( \tilde F_\pm(y))  =-\half \left (\Big (y^\ga \f{\p}{\p y^\ga}
 +2(1 \pm \nu ) \Big ) \tilde F_\pm(y) \mp\nu  ( \tilde F_\pm(-y_1,y_2)+
\tilde F_\pm(y_1,-y_2)) \right )\,.
\ee
Lorentz transformation has universal form in all cases
\be
L A(y) =\half \left (y^1 \f{\p}{\p y^1}- y^2 \f{\p}{\p y^2}
\right ) A(y)\,.
\ee

Essential difference between $AdS_3/CFT_2$ and $AdS_4/CFT_3$
dualities is that in the latter case 0-forms $C$ are glued to
the HS curvatures  at the linearized level by
Eq.~(\ref{CON12}), that leads to the nontrivial gluing of $3d$ conformal
HS currents to $3d$ conformal HS gauge fields via (\ref{d0}). In the
 $AdS_3$ HS gauge theory no gluing between HS gauge
fields and 0-forms $C$ occurs at the linear level. Moreover,
no nontrivial gluing of this type is even possible in $AdS_3$
 because $3d$ Chern-Simons HS gauge theory  admits no
Weyl tensor and its HS generalizations. This has a consequence that
$2d$ conformal fields $J$ associated with $C$ do not source $2d$
conformal HS gauge fields at the linearized level which, in fact,
allows conformal fields associated with $C$  to have continuous
conformal dimension parametrized by $\nu$.

This does not however imply that the $2d$ conformal fields
$J$ and $2d$ conformal HS gauge fields are completely independent.
$AdS_3$ HS gauge fields are sourced by the
HS currents built from bilinears of the $AdS_3$ matter fields
$C$ which represent the $3d$ stress tensor and its HS
generalizations. From the $CFT_2$ dual viewpoint this means
that $2d$ conformal HS curvatures will receive sources $T\sim JJ$
starting from the second order in $J$ where $T$
is a HS generalization of the stress tensor built
from the currents $J$.

To obtain  $z$-independent $2d$ equations from the $AdS_3$ HS theory
one should again properly rescale the oscillators similarly to
(\ref{w}) and (\ref{v}). To simplify formulae we abuse notations
denoting the rescaled variables by $y$.
Then $2d$ HS field equations have the following  structure in the
lowest order in which 0-forms contribute to r.h.s.s of the equations for
HS gauge fields
\be
\label{RT}
R(y,k,\psi|\xx,\tilde \xx) = d\xx d\tilde \xx
\f{\p^2}{\p y^1\p y^2} (T(y_1,y_2, k,\psi| \xx)+\widetilde
T(y_1,y_2,k,\psi|\tilde \xx))+\ldots \,,
\ee
where  the HS curvature is
\be
R(y,k,\psi|\xx,\tilde \xx) = d \Omega (y,k,\psi|\xx,\tilde \xx)+
\Omega (y,k,\psi|\xx,\tilde \xx)\star
\Omega (y,k,\psi|\xx,\tilde \xx)
\ee
with $\star$ denoting the non-commutative product of deformed
oscillators.
Equation (\ref{RT}) is to some extent analogous to Eq.~(\ref{dxzO})
if $T$ would be treated as an independent field. More precisely,
it is analogous to the $4d$ equations found in \cite{Gelfond:2010pm}
where current
interactions of $4d$ massless fields of all spins were constructed.

It remains to see  to which extent  the
scheme sketched above reproduces the $AdS_3/CFT_2$ HS duality conjectures of
\cite{Gaberdiel:2010pz,Chang:2011mz}. The construction of
conformal currents $T$ in terms of bilinears of the $2d$ fields $J$ resulting from
the $3d$ fields $C$ is analogous to Sugawara construction
considered in \cite{Chang:2011mz}. Correspondingly, the operator product
of  conformal conserved currents $T$ is expected to reproduce the
$W_\lambda$ algebra \cite{fig,Gaberdiel:2011wb} with
$
\lambda = \f{1\pm\nu}{2}\,.
$
Note that at the classical level the construction
of the nonlinear $\nu$--dependent $W$ algebra in terms of
HS algebras, which is anticipated to be equivalent to the
$W_\lambda$ algebras of  \cite{fig,Gaberdiel:2011wb},
was found  in \cite{Brink:1995uw}.
In any case, a conformal theory
 dual to the $AdS_3$ HS theory should be nonlinear.

\section{Higher-spin theory  and quantum mechanics}
\label{qm}

Unfolded dynamics provides a powerful direct
tool elucidating duality between theories in
various dimensions, sometimes going beyond the conventional framework
of $AdS_{d+1}/CFT_d$ duality
\cite{Maldacena:1997re,Gubser:1998bc,Witten:1998qj}. For instance,
one can consider a chain of $AdS_{n+1}/AdS_n$ dualities
as conjectured in \cite{BHS} (see also interesting recent work
\cite{Nilsson:2012ky}), using the chain of Poincar{\' e} foliations (\ref{fol}),
or, alternatively, by going directly from a higher dimension
to the lower one. An intriguing example of the latter
option considered in this section is provided by the duality
between HS theories in the matrix space $\M_M$, formulated
originally in the unfolded form in \cite{BHS}, and non-relativistic
quantum mechanics. This consideration is closely related to the
recent analysis of symmetries of quantum mechanical models in \cite{M,B,Bekaert:2011qd}.

Via appropriate rescaling and complexification of
variables, the rank-one equation (\ref{r1}) in $\M_M$
can be rewritten in the form
\be
\label{scheq}
\Big (i\hbar \frac{\partial}{\partial  X^{AB}} +
\f{\hbar^2}{2m}\frac{\partial^2}{\partial Y^A \partial Y^B}
\Big ) \Psi(Y|X)=0\q A,B=1,\ldots M\,.
\ee
(Note that the factor of $i$ in this equation naturally
appears in the analysis of HS equations in Siegel space
 \cite{gelcur}). As discussed in Section
\ref{globs}, maximal symmetries of the free unfolded equations
coincide with algebra $l^{max}(V)$ (of commutators) of endomorphisms
of the space $V$ where the 0-forms $\Psi(Y|X)$ at any $X=X_0$ are
valued. Hence, symmetries of the equations
(\ref{scheq}) are generated by various operators in the
space $F$ of functions of $Y^A$.

Generally, to specify a space of
operators in the functional space one has to specify their
properties in some more detail. To respect relativistic
symmetries, we should consider the space of differential
operators with polynomial coefficients in $Y^A$ which is
equivalent to the Fock space realization with the
``oscillators" $P_A$ and $Y^B$ that satisfy
\be
[P_A\,, Y^B] = \delta_A^B\q [P_A\,,P_B] =0\q [Y^A\,,Y^B] = 0\,.
\ee
In this terms, $F$ is the space of vectors $f(Y)|0\rangle$ induced
from the vacuum $|0\rangle$ satisfying
\be
P_A|0\rangle =0\,.
\ee
Hence, the symmetry algebra of Eq.~(\ref{scheq}) is
generated by various polynomials of $P_A$ and $Y^B$. This is the
generalized conformal HS algebra considered
in \cite{BHS}. It contains  $sp(2M)$ generated by
\be
K^{AB}= Y^A Y^B\q L^A{}_B = \{Y^A\,,P_B\}\q P_{AB} = P_A P_B\,.
\ee
The $X$-dependence of  global HS transformations determined
by Eq.~(\ref{glo0}) was  found in \cite{BHS}.

In \cite{Mar} it was shown that time-like directions in $\M_M$
are associated with positive-definite $X^{AB}$. In particular
one can set
\be
X^{AB}= t M \delta^{AB}\,,
\ee
where $t$ is the time evolution parameter. Restriction of Eq.~(\ref{scheq})
to $t$ gives usual $M$-dimensional Schrodinger equation
\be
\label{sch}
\Big (i\hbar \frac{\partial}{\partial  t} + \f{\hbar^2}{2m}
\delta^{AB} \frac{\partial^2}{\partial Y^A \partial Y^B}
\Big ) \Psi(Y|t)=0\,,
\ee
where $Y^A$ are now interpreted as coordinates of the
Galilean space.

{}From general properties of unfolded formulation discussed
in Section \ref{Unfolded dynamics}
it follows that relativistic rank-one equations in $\M_M$ are
equivalent to the nonrelativistic Schrodinger equation in $M$
dimension. The cases of $M=2$ and $M=4$ are particularly
interesting from the relativistic field theory perspective.
Eq.~(\ref{r1}) with $M=2$ describes
massless scalar ($\Psi(Y|X)=\Psi(-Y|X)$) and spinor
($\Psi(Y|X)=-\Psi(-Y|X)$) in 2+1 dimension. Eq.~(\ref{r1}) with $M=4$ describes
massless particles of all integer ($\Psi(Y|X)=\Psi(-Y|X)$) and
half-integer ($\Psi(Y|X)=-\Psi(-Y|X)$) spins in 3+1 dimension
\cite{BHS}.

It should be noted that relativistic systems in $\M_M$ are
conformal \cite{Mar,Bandos:2005mb}. In particular, $sp(4)$ is just the
$3d$ conformal algebra while $sp(8)$ contains
the $4d$ conformal algebra $su(2,2)$ as a subalgebra.
This immediately implies that these algebras do act on
solutions of the respective non-relativistic field equations
as well as the full Weyl algebra of operators built from $P_A$
and $Y^A$.
However this action does not look geometric in terms
of twistor variables $Y^A$ interpreted as space coordinates
of  nonrelativistic quantum mechanics.
(More precisely, beyond free field level, these are coordinates
$u^A$ introduced in Section \ref{ferext}.) Other way around,
nonrelativistic symmetries, which act geometrically in terms
of nonrelativistic coordinates $Y^A$, look nongeometric in terms
of relativistic coordinates.

This is manifestation of a very general situation. In the
unfolded dynamics approach it is  easy to introduce coordinates
in which any symmetry $h$ of a given system acts geometrically
by introducing an appropriate non-zero flat connection of $h$.
However different symmetries require different coordinates (spaces)
and connections. Description of the same system
in different space-times gives holographically dual theories.
Being obvious in  unfolded dynamics approach,
where it refers to the same twistor space (which is the space
of $Y^A$ in the quantum-mechanical model of interest), in other
approaches holographic duality may look obscure.

Eq.~(\ref{sch}) is Schrodinger equation for free
nonrelativistic particle. One may wonder what if the system
is deformed by a potential? In the framework of
unfolded dynamics, this does not affect the consideration
much, at least formally.
Indeed, in presence of potential $U(Y)$, the equation
\be
\label{schpot}
\Big (i\hbar \frac{\partial}{\partial  t} + \f{\hbar^2}{2m}
\delta^{AB} \frac{\partial^2}{\partial Y^A \partial Y^B}
-U(Y)\Big ) \Psi(Y|t)=0\,
\ee
remains linear, hence exhibiting infinite symmetries.
In the spirit of unfolded dynamics, it can
be interpreted as  flatness condition
\be
\label{Dsch}
D\Psi(Y|t)=0\q D = dt\f{\p}{\p t} + \Omega\q
\Omega = i\hbar^{-1} dt H\q H= -\f{\hbar^2}{2m}
\delta^{AB} \frac{\partial^2}{\partial Y^A \partial Y^B}
+U(Y)\,.
\ee
In the one-dimensional case with the single coordinate $t$,
any connection is flat, \ie the compatibility conditions for
Eq.~(\ref{Dsch}) are trivially satisfied. Hence it can be
represented in the pure gauge form which is simply
\be
\Omega=  \exp{(-i\hbar^{-1} H  t)}\, d\, \exp{(i\hbar^{-1} H t)}\,.
\ee
The same similarity transform relates symmetries of the $H=0$
system to those of $H\neq 0$.

Other way around, any HS connection $\Omega(Y|X)$ in $\M_M$
(not necessarily flat) generates a flat connection
$\Omega_t$ as its pullback to the time arrow. Hence any HS
geometry is holographically dual to some quantum mechanics.
For example, from Eq.~(\ref{tw}) we observe
that an appropriate $\lambda$-dependent rescaling maps
$AdS$ geometry to the harmonic potential
\be
U(Y) = \half m\go^2 Y^A Y^B \delta_{AB}\,,
\ee
where
the coupling constant is proportional the cosmological constant
$-\Lambda\sim \lambda^2$
\be
\half m\go^2 = \lambda^2\,.
\ee
On the other hand, $dS$ geometry is holographically dual
to the inverted harmonic potential with negative $\go^2$,
that is of course not too surprising in the context of
inflation.

The correspondence between relativistic
systems in higher dimensions and quantum mechanics is not
just formal. In particular, these holographically dual systems
have the same spectra. Namely, by virtue of unfolded equations,
the spectrum of states of free relativistic massless
particles of all spins in 3+1 dimension is identical to
that of four quantum harmonic oscillators, while the spectrum
of massless particles in 2+1 dimension is the same as of two
harmonic oscillators. Finite-dimensional Schrodinger algebra of
nonrelativistic symmetries of Schrodinger equation
(see e.g. \cite{Bekaert:2011qd} and references therein)
form a subalgebra of the algebra $sph(4|\mathbb{R})$ of Section
\ref{ferext}. In particular, so-called mass operator $\hat M$
 is represented
by the central element of the Heisenberg subalgebra of
$sph(4|\mathbb{R})$.

Let us note that the duality of relativistic and nonrelativistic
equations allows a natural interpretation for such a standard tool
for the study  of relativistic equations as oscillator
realization of relativistic symmetries extensively used for the
group-theoretic analysis of relativistic theories \cite{Bars:1982ep,Gunaydin:1998km}.
{}From the
holographic point of view pursued in this paper, it results from
the dual realization of the relativistic system in terms of its
nonrelativistic cousin. Moreover,
not only symmetries of holographically dual relativistic and
non-relativistic systems are the same. Their conserved
currents coincide as well. This is a
simple consequence of the analysis of \cite{gelcur} summarized
in Section \ref{currents}.
 The key fact is that a differential $2M$--form $\Omega$ (\ref{varpi})
 is closed in the correspondence space unifying
(relativistic) space-time $\M_M$ with coordinates $X^{AB}$ with the
twistor space ${\mathbf T}$ (non-relativistic space-time) with coordinates $Y^A$.
As a result, a conserved charge can be evaluated both in $\M_M$
and in ${\mathbf T}$. In the first case, it gives a relativistic conserved
charge in $\M_M$ while in the second case it appears as a
nonrelativistic conserved charge in ${\mathbf T}$. In fact, higher
 conserved charges of nonrelativistic quantum mechanics
 constructed recently in \cite{Bekaert:2011qd} just coincide with those
 resulting from the pullback of $\Omega$ to ${\mathbf T}$.

Surprisingly enough, equivalence of relativistic and non-relativistic
systems described above acquires interpretation of a Penrose transform
induced by the unfolded equations. This should have much in common with
the interpretation of non-relativistic physics as the relativistic one
in the light-cone higher-dimensional system
(see \cite{Bekaert:2011qd} and references therein).

\section{Towards off-shell formulation}
\label{actions}
The property underlying holographic duality
is that dynamics of universal unfolded systems
is characterized entirely by the differential $Q$ (\ref{qdif})
defined on the ``target space" of dynamical variables
independently of the original space-time. In particular,
invariants like actions and conserved
charges are characterized by  $Q$--cohomology \cite{act}.

First suppose that the system (\ref{unf}) is off-shell.
As shown in \cite{act}, a gauge invariant action
is an integral over a $d$-cycle $M^d$
\be \label{action1} S=\int_{M^d} \cl (W)\,.
\ee
of some $Q$-closed $d$--form {\it Lagrangian function} $\cl(W)$
\be
\label{qcl}
Q\cl=0\,:\qquad G^\ga (W)  \f{\p}{\p W^\ga} \cl(W) =0\,.
\ee
It is easy to see that, being $Q$--closed, $S$
 is invariant under the gauge transformations (\ref{delw}).
If $\cl$ is $Q$-exact, by virtue of (\ref{unf1})
it is $d$--exact, {\it i.e.} nontrivial invariant actions represent
$Q$ cohomology of the system in question.

If the system is on-shell and $\cl$ represents  $H^p(Q)$, the
same formula describes a conserved charge as
an integral over a $p$--cycle $\Sigma$
\be
q = \int_{\Sigma} \cl(W)\,.
\ee
Examples of application of this construction were given in \cite{act}.
Let us stress that, the analysis in terms of $Q$--cohomology
applies to both linear and nonlinear unfolded systems.

In unfolded dynamics, Noether current interactions
are directly related to conserved currents. In the case of interest
they result from the expression for the
conserved charge (\ref{Q}), (\ref{eta_f}). For example,
in the case of $M=2$, since the
4-form $\Omega$ (\ref{Q}) is closed for any $\tilde T_\eta$
(\ref{eta_f}) with $\eta(\W,\,Y|x)$ that satisfies
Eq.~(\ref{unfol2_Fur}), the 5-form
\bee
L &&\ls\!\!=  \Omega (\W,\,Y\,|X) d\,\W_\gb \,d\,W^\gb
\Big(i\,\W{}_\gb{} d\, X{}^{\ga\gb} -    d\,Y{}^\ga
\Big) (i\,\W{}^\gga{} d\, X{}^{\gga}{}_\ga -    d\,Y{}_\ga
\Big)\nn\\
&&\int d\,U_\gga \,d\,U^\gga
\exp{-i W_\ga
U^\ga}
\,\, J(U,\,Y\,|X)\,
\eee
is closed up to $J^2$ terms by virtue of the unfolded equations
(\ref{dO}).

Let us look more closely at the relation between on-shell and off-shell
systems. Let $W_{on}^\Omega$ be a set of forms
of some on-shell system. Its off-shell extension should
contain additional fields $E^a$ that appear on the r.h.s.s
of the field equations to replace the differential equations
 by constraints expressing new fields
via l.h.s.s of the field equations. Abusing notation we
can write
\be
\label{eq}
L^i(W(x))= \ce^i(x)\,,
\ee
where $L^i(W(x))$ describes l.h.s.s of the dynamical
equations on $W^\Omega$. Note that $\ce^i$ is a part of the full set
of $E^a$ since, in  unfolded dynamics, $E^a$
contains $\ce^i$ along with all their derivatives. We will call
$\ce^i$ {\it primary off-shell fields}, saying that
they glue the field equations of the on-shell
system in question. This is equivalent to the statement that
primary off-shell fields match the on-shell
$\sigma_-$ cohomology associated with the l.h.s.s of the
field equations to enforce  the corresponding cohomology
of the off-shell system be zero.

The off-shell system with $W_{off}^A=(W^\Omega_{on}, E^a)$ is
such that  $E^a=0$ puts the off-shell system on shell.
This means that the off-shell system is described by such
$G^A(W_{off})$ that both
\be
Q^{off} =G^A(W_{off})\f{\p}{\p W^A_{off}} =
G^\a(W_{on},E)\f{\p}{\p W^\a_{on}} + G^a(W_{on},E)\f{\p}{\p E^a}
\ee
and
\be
Q^{on} = G^\a(W_{on},0)\f{\p}{\p W^\a_{on}}
\ee
are nilpotent
\be
Q^{off}Q^{off}=Q^{on}Q^{on} =0\,.
\ee
This is  a consequence of  the property
\be
\label{pro}
G^a(W_{on},0)=0\,,
\ee
which should be true for any  off-shell extension.
Indeed, otherwise,
the on-shell fields $W^\a_{on}$ would source the fields $E^a$
 not allowing to put the system on shell.
Field equations imply $\E^i=0$ and hence
\be
E^a =0\,.
\ee

To extend the on-shell analysis of this paper to the
full quantum level  an off-shell extension of
the system has to be considered. This problem has not been yet solved in a fully satisfactory
way. An interesting hint from the
analysis of  \cite{Gelfond:2010pm} is that the system, that
describes current interactions of $4d$ massless fields, can be
viewed  as the $4d$ off-shell system with the current fields
$J$  interpreted as off-shell fields $\ce^i(x)$. On the other hand, the same fields
can be interpreted either as describing two-particle states in the
system or as free $6d$ fields. This suggests the idea that
proper account of off-shell quantum effects
in terms of unfolded dynamics  may result from consideration of the
theory in higher and higher dimensions, allowing
to interpret quantum-mechanical effects as classical dynamics
in an infinite-dimensional space that has enough room to describe
all multiparticle states of the  system.

In a more traditional fashion, if an off-shell action
is available in the unfolded formulation, it can be
used to produce generating functionals in the standard
path integral approach. Again, the idea is that using that
an action functional is closed in an appropriate correspondence
space which extends space-time with some twistor coordinates,
the integration can be  performed
in the twistor space for all holographically dual theories.
In that case, various holographic interpretations of
the same generating functional will be fully equivalent.

Actions of the form (\ref{action1}) are also appropriate for the
analysis of anomalies in the formulation in the doubled bulk space
suggested in Section \ref{bound}. Although naive interpretation of the
action (\ref{action1}) may be ill-defined because of divergencies at
$z=0$ it can be regularized via deformation of the
integration contour to the complex plane in $z$, say, via  substitution
$z\to z +i\epsilon$. Anomalous terms will be associated with
 singularities in $\epsilon$. In fact,  complexification of matrix
coordinates $X^{AB}$ in HS theories has been used in \cite{gelcur}
to regularize integrals for HS conserved charges analogous to the action integral
(\ref{action1}) where $X^{AB}$ were complexified to ${\mathcal Z}^{AB}$
from the upper Siegel
half-space. In the example of \cite{gelcur} it was shown that the charge
integrals are independent of variations of a complex integration contour away from
singularity. If the same happens in HS theories this would imply that the
regularized action is independent of $\epsilon$ hence being anomaly free
(though it may be dependent on the  contour homotopy class). The same time,
independence of local contour variation implies $Q$-closure of the action
(\ref{action1}) which, in turn, implies its gauge invariance \cite{act}.
It would be  interesting to see how this scheme works in off-shell HS theories.

Note that for HS theories formulated  in  matrix space
$M_\M$, regularization via deformation to Siegel space has
deep meaning in various respects \cite{gelcur}. In particular,  solutions analytic in upper and
lower Siegel spaces correspond, respectively, to particles and antiparticles. Also solutions
of HS equations from the upper Siegel half-space, that are periodic in the $Y^A$,
are closely related to Riemann theta-functions where
complexified coordinates ${\mathcal Z}^{AB}$ acquire the meaning of a period matrix.

\section{Conclusion}
\label{conc}

In this paper it is demonstrated how holographic duality results
from different interpretations of one and the same theory.
This phenomenon is very general  and  applies
to any theory (not necessarily conformal).
To establish  holographic duality it is most useful to
reformulate a theory in the unfolded form \cite{Ann} of
coordinate independent first-order equations formulated
in terms of exterior differential as space-time derivative
and differential forms as field variables. Once such a
formulation is achieved, one can play freely with
space-time dimension, adding or removing coordinates without changing
dynamical content of the theory. This provides a vast variety of
differently looking models in space-times of different dimensions
which however are by construction locally equivalent.
Since unfolding machinery applies to any theory,
every  model belongs to a class of holographically
equivalent models.

Since HS theories were originally formulated
within the unfolded dynamics approach, they provide a
 natural arena illustrating this phenomenon. In this paper
 we focused on the $AdS_4/CFT_3$ and
$AdS_3/CFT_2$ HS dualities. The latter was put forward in
\cite{Henneaux:2010xg,Campoleoni:2010zq,Gaberdiel:2010pz,Chang:2011mz}.
The former was conjectured by Klebanov and Polyakov
 \cite{Klebanov:2002ja} to relate the simplest
 $AdS_4$ HS theory to $3d$ $O(N)$ sigma-model and was
 partially proved by  Giombi and Yin \cite{Giombi:2009wh,Giombi:2010vg}
  for correlators involving any three spins
$s_1, s_2, s_3$ that do not respect the triangle inequality.

Recently, Maldacena and Zhiboedov conjectured \cite{Maldacena:2011jn}
 that $AdS_4$ HS theory is  dual to the $3d$ free model
even beyond the large $N$ limit.
The arguments of \cite{{Maldacena:2011jn}} are very general, generalizing
the Coleman-Mandula theorem \cite{cm} to conformal theories.
Namely, the authors of \cite{{Maldacena:2011jn}} have shown that if a
unitary local conformal field theory possesses a conserved
HS current then it must be a theory of currents of free conformal
fields. Since the $AdS_4$ HS theory possesses HS symmetries, the
conclusion of \cite{Maldacena:2011jn} was that its boundary dual is free.

Analysis of this paper shows however that, except for two particular cases,
the boundary theory dual to $AdS_4$ HS theory
turns out to be nonlinear, escaping some of  conditions of the
Maldacena-Zhiboedov theorem. Namely, the boundary theory describes
interactions of conformal currents in the framework of  $3d$ conformal HS
gauge theory  which extends $3d$
(Chern-Simons) conformal gravity to higher spins.
Being a gauge theory, it is not unitary, while a particular
gauge choice makes it nonlocal and/or not conformal.
Another property of the boundary dual of the $AdS_4$ HS theory
is that boundary conformal currents associated with massless
fields in $AdS_4$ are not conserved in the usual sense being instead
covariantly conserved with respect to the  $3d$ conformal HS algebra.
Analogous phenomena are expected to take place for higher dimensions
$d>4$, relating nonlinear HS theories in any $d$ \cite{Vasiliev:2003ev}
to boundary conformal HS theories in $d-1$. However, this  duality is
expected to be far more complicated
because of complicated structure of the corresponding generalized
twistor space.

We have identified two particular truncations of the bulk HS theories
which have free bosonic and fermionic boundary duals in agreement with
the conjectures of Klebanov and Polyakov \cite{Klebanov:2002ja} and
Sezgin and Sundell \cite{Sezgin:2003pt}. In this cases $3d$ conformal HS
gauge fields decouple from the boundary currents and the corresponding
boundary theories indeed turn out to be free in agreement with the
Maldacena-Zhiboedov theorem \cite{Maldacena:2011jn}. Truncations to the free
boundary theories are based on the parity automorphism $P$ of the $AdS_4$ system
that reflects the Poincar{\'e} coordinate $z$. Its application requires the
doubling of the Poincar{\'e} chart, identifying the $AdS_4$ boundary with
the stationary surface of $P$.

In the setup with doubled bulk space, it is not necessary to impose
definite boundary
conditions at $z=0$ since it becomes a regular point in terms of appropriately
rescaled twistor variables. Hence, in our approach, the $3d$ dual  of
$AdS_4$ HS gauge theory describes a doubled number of $3d$ currents which in particular
contain two scalar currents of different dimensions. Generally, all these
currents interact via $3d$ conformal HS gauge fields.
We believe that the trick with doubled bulk in $AdS/CFT$, which follows naturally
from unfolded dynamics, should also have interesting applications  beyond HS theories.

A new phenomenon found in this paper is that both of holographically
dual theories  are theories of (conformal) gravity. This phenomenon
seems to be very general and should take place beyond the $N\to\infty$ limit
for most of holographic models of bulk gravity as a consequence of
coordinate independence of the unfolded formulation.

In this paper we did not check explicitly how our prescription
reproduces conformal correlators on the conformal side, leaving
discussion of this issue to the future work \cite{GC}.

Taking into account that non-Abelian
contributions to conformal HS curvatures exist only for spins
$s_1, s_2, s_3$ that respect the triangle inequality,
it would be interesting to see whether nonlinear corrections
of the boundary theory can help to conform the boundary and bulk
calculations of \cite{Giombi:2009wh} for such spins.

Analysis of this paper is mostly on-shell, operating
in terms of field equations rather than action since
 HS actions are not yet
known to all orders on the both sides of the HS duality.
Once they are available, the analysis
can be immediately extended to the  action level.
Hence, most urgent problems for the future  include explicit
construction of nonlinear $3d$ HS conformal gravity and action
functionals for both  $AdS_4$ HS theory and  its $3d$ dual.

A peculiar feature emerged from the analysis of
the particular HS model in this paper is that
the infinite boundary limit is not a necessary ingredient of the
duality which can formally
be established on every co-dimension one surface $\Sigma$ in the
bulk. However, for general $\Sigma$,
the relation between fields and sources in the dual theories,
that respects conformal symmetry, is nonlocal
while in the infinite boundary limit $z\to 0$
the relation turns out to be local in accordance with the
standard prescription of \cite{Gubser:1998bc,Witten:1998qj}.
Being complicated in terms of space-time coordinates,
the nonlocal holographic duality map between two theories
on general $\Sigma$ acquires natural meaning in terms of
non-commutative twistor variables, describing the map
between  Weyl and Wick star products.
It should be noted however that transition from one ordering
prescription to another may, in principle, lead to divergencies in the
star-product formalism in HS theories because the construction
involves nonpolynomial elements like Klein operators
(\ref{kk4}). When this happens, a model exhibits conformal anomaly.

Systematic reformulation of unfolded theories in terms of twistor
variables greatly simplifies analysis of holographic duality making it
nearly tautological. Seemingly different theories are described by
solutions of the same equations
in the generalized twistor space or by the same action-like invariants
evaluated as integrals over  twistor variables. Two holographically
dual models result from different space-time extensions of the same
twistor model.

In \cite{BHS} it was conjectured that massless conformal
HS theories may form a chain of dualities between models in
space-times of different dimensions. If a
 boundary theory contains conformal gravity
it can be again put in locally $AdS_d$
background, say, by using the foliation prescription of
Section \ref{conff}.
In the end,  one stops at some $2d$ conformal theory, $1d$
quantum-mechanical theory or even $0d$ matrix-like theory, which
is nothing but the part of the theory reduced solely to the
twistor space (e.g., equations (\ref{SB}) and (\ref{SS}) in the
$AdS_4$ HS system).

Duality between HS theories and nonrelativistic
quantum mechanics discussed in Section \ref{qm} provides an
exciting example of $1d$ dual interpretation.
Deep relation between HS theories and quantum mechanics
makes it difficult to refrain from speculation that the
two  systems may be literally equivalent while their
different interpretations depend on particular details of
physical observation in question. In other words, it is tempting to rise a
risky question whether HS theories can tell us what quantum
mechanics is. This issue has too many aspects to be discussed
in detail in this paper. However, one immediate consequence
 is that, if true, nonlinear HS theories should
imply that Schrodinger equation has to receive nonlinear
corrections of the form prescribed by HS theory. Since the
coupling constant inherited from HS theories should be related
to the gravitational constant, nonlinear
corrections to quantum mechanics should be negligible in the
non-relativistic regime. Nevertheless one can speculate that
their appearance
may shed some light on such conceptual problems of quantum mechanics
 as, for instance, momentary wave packet reduction.

Tremendous robustness of the quantum gravity problem suggests that its
solution may require modification of the both ingredients.
HS theory may provide a framework for nontrivial
merge of gravity with quantum mechanics, affecting the present-day
understanding of both. If so, non-relativistic quantum
mechanics may one day provide us  with an unexpected tool for
the study of quantum gravity in  laboratory experiments.
At any rate, we believe that HS
gauge theory has  potential to unify gravity and quantum
mechanics in a nontrivial and constructive way.

\section*{Acknowledgments}
I am grateful to O.Gelfond and E.Skvortsov for stimulating discussions
and to E.Joung, V.Losyakov, R.Metsaev, P.Sundell and, especially, V.Didenko
for useful comments. This research was supported
in part by RFBR Grant No 11-02-00814-a and Alexander von Humboldt
Foundation Grant No PHYS0167.

\end{document}